\documentclass[fleqn]{aa}
\usepackage{graphicx}
\usepackage{natbib}
\usepackage{color}
\usepackage{fancyvrb} 
\usepackage{amssymb}
\usepackage{amsmath}
\usepackage[colorlinks,allcolors=blue]{hyperref}
\usepackage{bm,url,times}
\graphicspath{{../notes/onlyplots/fig2/}}

\newcommand{\Eq}[1]{Eq.~(\ref{#1})}
\newcommand{\Fig}[1]{Fig.~\ref{#1}}
\newcommand{\Tab}[1]{Table~\ref{#1}}
\newcommand{\dd}{{\rm d} {}}
\newcommand{\EQ}{\begin{equation}}
\newcommand{\EN}{\end{equation}}
\newcommand{\EQA}{\begin{eqnarray}}
\newcommand{\ENA}{\end{eqnarray}}
\newcommand{\Eqs}[2]{Eqs.~(\ref{#1}) and~(\ref{#2})}
\newcommand{\con}{\Omega_{\star}}
\newcommand{\meanUU}{\overline{\mbox{\boldmath $U$}}{}}{}
\newcommand{\mean}[1]{\overline #1}
\newcommand{\nab}{\mbox{\boldmath $\nabla$} {}}
\newcommand{\NNNN}{\mbox{\boldmath ${\sf N}$} {}}
\def\urms{u_{\rm rms}}
\def\nut{\nu_{\rm t}}
\newcommand{\PC}{{\sc Pencil Code}~}
\newcommand{\UU}{\mbox{\boldmath $U$} {}}
\newcommand{\SSS}{\mbox{\boldmath $S$} {}}
\newcommand{\Sec}[1]{Section~\ref{#1}}
\newcommand{\qyz}{Q_{yz}}
\newcommand{\qxz}{Q_{xz}}
\newcommand{\qxy}{Q_{xy}}

\newcommand{\qzz}{Q_{zz}}

\newcommand{\tqyz}{\widetilde{Q}_{yz}}
\newcommand{\tqxz}{\widetilde{Q}_{xz}}
\newcommand{\tqxy}{\widetilde{Q}_{xy}}
\newcommand{\tqxx}{\widetilde{Q}_{xx}}
\newcommand{\tqyy}{\widetilde{Q}_{yy}}
\newcommand{\tqzz}{\widetilde{Q}_{zz}}

\newcommand{\mux}{\overline{\mbox{\boldmath $U$}}{}_x}{}
\newcommand{\muy}{\overline{\mbox{\boldmath $U$}}{}_y}{}
\newcommand{\muz}{\overline{\mbox{\boldmath $U$}}{}_z}{}
\newcommand{\tmux}{\widetilde{\UU}_x}
\newcommand{\tmuy}{\widetilde{\UU}_y}
\newcommand{\tmuz}{\widetilde{\UU}_z}
\newcommand{\meanrho}{\overline{\rho}}
\newcommand{\nupar}{\nu_{\parallel}}
\newcommand{\tnupar}{\widetilde{\nu}_{\parallel}}
\newcommand{\nuper}{\nu_{\perp}}

\newcommand{\tqxzl}{\tqxz^{(\Lambda)}}
\newcommand{\tqxzn}{\tqxz^{(\nu)}}

\newcommand{\qyzl}{\qyz^{(\Lambda)}}
\newcommand{\qyzn}{\qyz^{(\nu)}}

\newcommand{\qxyl}{\qxy^{(\Lambda)}}
\newcommand{\qxyn}{\qxy^{(\nu)}}

\title{Generation of mean flows in rotating anisotropic turbulence:
The case of solar near--surface layer}

\author{A. Barekat\inst{1} \and M. J. K\"apyl\"a\inst{2,1,3},
  P. J. K\"apyl\"a\inst{4}, E. P. Gilson\inst{5} \and H. Ji\inst{5,6}}  

\date{\today}

\institute{
Max-Planck-Institut f\"ur Sonnensystemforschung, Justus-von-Liebig-Weg
3, 37077 G\"ottingen, Germany
\and
Department of Computer Science, Aalto
University, PO Box 15400, FI-00076 Aalto, Finland 
\and
Nordita, KTH Royal Institute of Technology and Stockholm University,
Roslagstullsbacken 23, SE-10691 Stockholm, Sweden
\and
Institut f\"ur Astrophysik, Georg-August-Universit\"at G\"ottingen,
37077 G\"ottingen, Germany
\and
Department of Astrophysical Sciences, Princeton University, Princeton,
New Jersey 08544, USA
\and
Princeton Plasma Physics Laboratory, Princeton University, Princeton,
New Jersey 08543, USA
}
\begin{document}

\abstract
{
Helioseismic results indicate that the radial gradient of the rotation 
rate in the near--surface shear layer (NSSL) of the Sun is independent 
of latitude and radius.
Theoretical models utilizing the mean--field approach have been
successful in explaining this property of the NSSL, while global
direct or large--eddy magnetoconvection models have so far been unable
to reproduce it.
}
{
We investigate the reason for this discrepancy by measuring the mean
flows, Reynolds stress, and turbulent transport coefficients under
conditions mimicking those in the solar NSSL. 
}
{
Simulations with minimal ingredients for the generation of mean flows are
studied. These ingredients are inhomogeneity due to boundaries, anisotropic
turbulence, and rotation. Parameters of the simulations are chosen such
that they match the weakly rotationally constrained NSSL.
The simulations probe locally Cartesian patches of the star at a given
depth and latitude. The depth of the patch is varied by changing the 
rotation rate such that the resulting Coriolis numbers cover the same range
as in the NSSL.
We measure the turbulent transport coefficient relevant for the 
non--diffusive ($\Lambda$--effect) and diffusive (turbulent viscosity) parts of
the Reynolds stress and compare them with predictions of current
mean--field theories. 
}
{
A negative radial gradient of mean flow is generated only at the equator where
meridional flows are absent. At other latitudes the meridional flow is
comparable to the mean flow corresponding to differential rotation. We
also find that meridional components of the Reynolds stress cannot be
ignored. 
Additionally, we find that the turbulent viscosity is quenched by
rotation by about $50$\% from the surface to the bottom of the NSSL.
}
{
Our local simulations do not validate the explanation for the generation of 
the NSSL from mean--field theory where meridional flows and stresses are 
neglected. 
However, the rotational dependence of turbulent viscosity in our simulations is 
in good agreement with theoretical prediction.
Moreover, our results are 
in qualitative agreement with global convection simulations in that a NSSL can
only be obtained near the equator.
}


\maketitle
\section{Introduction}
\label{sec:Intro}
The convection zone (CZ) of the Sun, despite being highly turbulent,
shows a well--organized large--scale axisymmetric rotation profile
depending on both depth and latitude. The entire CZ rotates faster at
the equator than at the poles and the rotation rate decreases mildly
with depth except near the radial boundaries where there are regions
of strong shear \citep{MT96,JS98}. Additionally, a large--scale  
circulation in the meridional plane, known as the meridional
flow (MC), is also present. The amplitude of MC is about 15-20
ms$^{-1}$ which is two orders of magnitude smaller than the rotational
velocity \citep{TD:79,DH96}.   

The near--surface shear layer (NSSL) occupies about 17\% of the CZ, or 
roughly $35\rm Mm$ in depth, from the photosphere. Recently, two 
further properties of it have been reported. 
First, the value of the logarithmic radial gradient of the rotation
rate is reported to be 
\EQ
\frac{\dd\ln\Omega}{\dd\ln r} \approx -1 
\EN
 in the upper 13 Mm of the NSSL independent of latitude up to
 $60^\circ$ \citep{BSG14}. Second, the gradient is evolving over time, 
by an amount between 5--10\% of its time--averaged value, following
closely the magnetic activity cycle \citep{BSG16}. 
On the other hand, the MC maintains its poleward motions throughout
the cycle \citep{HU14}.

Shear flows play an important role in generating and maintaining the
solar magnetic field and its activity cycle
\citep[e.g.][]{1980opp..bookR....K}. In particular, radial shear is
important in the $\alpha\Omega$ dynamo model for explaining the
equatorward migration of the magnetic activity \citep{EP95,HY75}. In
this model, negative radial shear in combination with positive
$\alpha$ is required to produce the correct equatorward migration of
the activity. Such negative shear exists only in the NSSL in the solar
CZ. The effect of the NSSL has been tested numerically in mean--field
dynamo models by \cite{KKT06} where it was found to aid equatorward
migration. More observational and theoretical arguments for the NSSL
strongly shaping the solar dynamo process were presented by
\cite{AB05}. The role of NSSL can be easily investigated in
mean--field models where it can be added or removed by hand. On the
contrary, global 3D convection simulations typically fail in
generating a realistic NSSL self--consistently
\citep[e.g.][]{2013ApJ...779..176G,HRY15} and thus its role on the
resulting dynamo solutions is unclear. Therefore, understanding the
role that the NSSL plays for the dynamo requires that we first
understand its formation mechanism and why global simulations do not
capture it. 

The equations governing the generation of large--scale flows in the
solar CZ are the following: First, azimuthally averaged angular
momentum equation describes the time evolution of the differential
rotation.  
This equation is obtained using the Reynolds decomposition,
where each physical quantity, $A$, is decomposed into its mean
$\mean A$ and fluctuations around the mean, $a$, and
where averages are taken over the azimuthal direction.
Then, we obtain the equation
\EQA
\frac{\partial }{\partial t}(\mean \rho \varpi ^2 \Omega)=&-&
\bm\nabla\bm\cdot \{ \varpi[\varpi \overline{\rho {\bm U}^m}\ \Omega +  \mean
\rho Q_{\phi i}-2\nu \mean \rho \overline{\SSS}\bm\cdot
\hat{\bm\phi} \nonumber\\
&-& (\overline{B}_\phi\overline{\bm B} / \mu_0 +M_{\phi i} )] \},
\label{eq:AM}
\ENA
where $\mean\rho$, $\overline{\bm
  U}^m=(\overline{U}_r,\overline{U}_\theta,0)$,
$\Omega=\overline{U}_\phi/r \sin\theta$, $\nu$, $\mu_0$, and
$\overline{\bm B}$ are density, meridional flow, angular velocity,
molecular viscosity, the vacuum permeability, and the magnetic field,
respectively. Furthermore, $\varpi=r\sin\theta$, where $\theta$ is the  
latitude, $Q_{\phi i}$ and $M_{\phi i}$ are the Reynolds and Maxwell
stresses, and $\overline{\SSS}$ is the mean rate of strain tensor. The
Reynolds and Maxwell stresses are the correlations of fluctuating
components $Q_{\phi j}=\overline{u_{\phi}u_j}$ and $M_{\phi
  j}=\overline{b_{\phi}b_j}/\mu_0$, respectively. Density fluctuations
are omitted corresponding to an anelastic approximation. 

Second, the azimuthally averaged equation for the azimuthal component
of vorticity, describes the time evolution of MC:  
\EQA
\frac{\partial \mean w_{\phi}}{\partial t}&=&\varpi \frac{\partial
  \Omega^2}{\partial z}\!+\!
(\bm\nabla\mean s \times \bm\nabla \mean T)_{\phi}
\!-\!\left[ \bm\nabla \!\times\!
  \frac{1}{\mean \rho}[\bm\nabla\!\bm\cdot\!(\mean \rho {\bm Q}\!-\!2\nu\mean \rho
    \overline{\SSS})] \right]_{\phi} \nonumber \\
&+&[\bm\nabla\times\bm\nabla\bm\cdot(\overline{{\bm B}{\bm B}}^T+{\bm M})],
\label{eq:MC}
\ENA
where $ \overline{\bm w}=\nabla \times \overline{\bm{U}}$ is the
vorticity, $s$ is the specific entropy, $T$ is the temperature, and
$\partial/\partial z$ is the derivative along the rotation axis. The
first and second terms describe the centrifugal and baroclinic
effects, respectively. From these two equations it becomes clear that
meridional flow can drive differential rotation, and vice versa, and
additionally any misalignment of density and temperature gradients can
drive meridional circulation through the baroclinic term, while
turbulent stresses are important in driving both flows. 

Theoretical studies have shown that the major players generating stellar 
differential rotation are the first two terms in both \Eqs{eq:AM}{eq:MC}
\citep{GR89,LK13}.
Additionally, the Coriolis number $\con$, describing the degree of 
rotational influence on the flow, defined as
\EQ
\con=2\tau\mean{\Omega},
\label{eq:con}
\EN
where $\mean{\Omega}$ is the rotation rate of the star and $\tau$ is
the turnover time of the turbulence, has been found to be a key
parameter. It describes the role rotation plays in different parts of
the CZ, in particular leading to a completely different rotation
profile within the NSSL in comparison to the rest of the CZ. 

In the solar structure model of \cite{Stix:2000}, $\con$ changes from
the surface to the bottom of the CZ as $10^{-3}\lesssim\con^{\rm
  NSSL}\lesssim 1 \lesssim\con^{\rm CZ}\lesssim 10$. Non--rotating
density--stratified convection is dominated by vertical motions in
which case the vertical anisotropy parameter $A_{\rm V} \propto u_{\rm
  H}^2 - u_r^2 < 0$, where $u_{\rm H}$ and $u_r$ are the turbulent
horizontal and radial velocities. Rotation tends to suppress
convection \citep[e.g.][]{Ch61} and typically $A_{\rm V}$ decreases  
when $\Omega_\star$ increases such that the maximum of $A_{\rm V}$ is
achieved for $\Omega_\star = 0$ \citep[e.g.][]{Chan2001,KKT04}. On the
other hand, rotation introduces horizontal anisotropy $A_{\rm H}
\propto u_\phi^2 - u_\theta^2$, where $u_\phi$ and $u_\theta$ are the
longitudinal and latitudinal velocities. Typically $A_{\rm H}$ is  
positive and it is increasing with $\Omega_\star$. Furthermore,
$A_{\rm H} \rightarrow 0$ as $\Omega_\star \rightarrow 0$.
Thus, $\con$ in the solar CZ reflects also the anisotropy of 
turbulence which arises due to the presence of the Coriolis force and
density stratification. Consequently, rotation and gravity vectors 
define the necessary two misaligned preferred directions for non--zero 
off--diagonal Reynolds stress \citep{GR89}.

A theoretical model that reproduces the entire rotation profile of the
Sun including the NSSL was presented in \citet[][hereafter
  KR05]{LKGR05}. They utilized a hydrodynamic mean--field (MF) model,
considering the properties of the turbulent flow explained above and
parameterized the Reynolds stresses in the form of turbulent transport
coefficients \citep[][see also \Sec{sec:theory}]{GR80,GR89}. 
They obtain the NSSL by taking the anisotropy of turbulence near the
surface into account such that $A_{\rm V}\gg A_{\rm H}$ for
$\con\lesssim 1$. This leads to strong inward transport of the angular
momentum in the NSSL and ultimately to the generation of the radial shear.
The remarkable agreement of recent observed latitudinal independence
of the gradient with their model brought motivation to develop the
theory further including the effect of the magnetic field in the NSSL
\citep{LK16}. This leads to the prediction of the time variation of
the angular velocity gradient during the solar cycle, qualitatively
agreeing with the observations. As the variations are caused by the
magnetic field, \cite{LK16} suggested that measurements of the
rotational properties of the NSSL can be used as an indirect probe for
measuring the sub--surface magnetic field. In their model, however,
the Reynolds stresses were computed using second--order correlation
approximation (SOCA), the validity of which in astrophysical regimes
with high Reynolds numbers is questionable.

To avoid the necessity of using such simplifications, it is desirable
to build numerical simulations of stellar convection, directly solving
for the relevant, either hydro-- or magnetohydrodynamic, equations in
spherical geometry. Such models have been developed and utilized since
the 1970s \citep[e.g.][]{Gi77,Gi83,Gl85}, but reproducing the NSSL has
turned out to be a serious challenge for these models. With such
global convection simulations it is possible to generate a shear layer
close to the equator, mostly confined outside the tangent cylinder,
where rotation--aligned, elongated large--scale convection cells form
\citep[see e.g.][]{RC01,KMB11,GSdGDPKM16,WKKB16,MT19}. Only when
higher density stratification has been used, a shear layer extending
to $60^\circ$ latitudes has been found \cite{HRY15}. In this case,
however, the gradient of the radial shear was positive in the range
$0^\circ<\theta<45^\circ$, contrary to the helioseismic inferences of
the NSSL. They concluded that the meridional Reynolds stress,
originating from the radial gradient of the poleward meridional flow,
is the most important driver of the NSSL. In their model, the
luminosity was decreased to obtain an accelerated equator, hence the
influence of rotation on convection (Coriolis number) was
overestimated, and they also speculated about an unfavourable
influence of boundary conditions to their results. Hence, it is
unclear whether these results really are applicable to the
NSSL. Overall, using global direct numerical simulations (GDNS) to
study the origin of the NSSL is cumbersome due to high computational
cost, the multitude of effects present, and the difficulty to reliably
separate them from each other. For such an approach, a simpler
modelling strategy is required which is attempted in this paper.

In addition to MF and GDNS models, the NSSL has also been studied from
the point--of--view of different types of equilibria. The most recent
of them, \cite{GB19}, considers the formation of the NSSL in a
magnetohydrostatic equilibrium model, being driven by a poleward
meridional flow near the surface. In addition to the assumption of
stationarity, while the magnetic field of the Sun is oscillatory, the
model considers only non--turbulent states; nevertheless, a
large--scale poloidal flow, when inserted on top of the equilibrium
configuration, is seen to reduce the rotational velocity near the
surface, hence leading to NSSL--like condition there.  
In the study of \cite{MH11}, the Reynolds and Maxwell stresses were
accounted for in the governing equations, hence allowing for turbulent
effects. They considered a case, where an equilibrium condition exists
for the angular momentum transport, Eq.~(\ref{eq:AM}), in which case
the meridional circulation and the relevant stresses must balance. Any
imbalance in the term encompassing the stresses was then postulated
not only to drive differential rotation, but more importantly to
induce a meridional flow. Similarly, the azimuthal vorticity equation,
Eq.~(\ref{eq:MC}), in a steady state, was postulated not only to drive
meridional flow, but more importantly contribute to maintaining the
differential rotation profile. In the earliest scenario explaining the
NSSL, \cite{FJ75} proposed that the reason for the existence of it
would be the local angular momentum conservation from rising and
falling convective fluid parcels, which would lead to inward angular
momentum transport. In the scenario of \cite{MH11}, however, such
angular momentum transport is not a sufficient condition to sustain
the NSSL, but another necessary ingredient is the meridional force
balance in between the turbulent stresses and centrifugally--driven
circulation within the NSSL. In the bulk of the convection zone,
meridional force balance would be rather provided by the baroclinic
effect, and the bottom of the NSSL would be determined by the
transition point from baroclinic to Reynolds stress balancing. Some
agreement with this scenario was found in the study by \cite{HRY15},
whose models showed that in the region of NSSL, the force caused by
the turbulent stresses was balanced by the Coriolis force. 

In this paper we adopt an entirely different approach to those
reviewed above. We formulate a model with minimal ingredients for the
generation of large--scale flows to study the role of
rotation--induced Reynolds stress specifically in a rotational regime
relevant for the NSSL. This involves replacing convection with
anisotropically forced turbulence and omitting density stratification,
magnetic fields, and spherical geometry. The simplicity of the model
allows unambiguous identification of the drivers of mean flows which
can be used to assess the generation mechanisms of the solar NSSL.

\section{The NSSL in terms of mean--field hydrodynamics}
\label{sec:theory}
In this section we briefly explain the theory of the $\Lambda$--effect
and its relevance for formation of the NSSL \citep{LK13,LK16}. We
refer the reader to \cite{GR89} for a thorough treatise. In this
theory, rotating and anisotropic turbulence contributes to diffusive
and non--diffusive transport of angular momentum. The non--diffusive
part is known as the $\Lambda$--effect \citep{Leb41}. Therefore, the
Reynolds stress consists of two parts   
\EQA
Q_{ij}&=&Q_{ij}^{(\nu)}+Q_{ij}^{(\Lambda)},\label{eq:R1}\\
Q_{ij}&=&N_{ijkl}\meanUU_{k,l}+\Lambda_{ijk}\Omega_k,\label{eq:R2}
\ENA
where $N_{ijkl}$ and $\Lambda_{ijk}$ are fourth-- and third--rank
tensors describing the turbulent viscosity and $\Lambda$--effect,
respectively. In spherical geometry $Q_{r\phi}$, $Q_{\theta \phi}$,
and $Q_{r\theta}$ are the vertical, horizontal and meridional
stresses, respectively. We note here that the meridional stresses
appear only in the vorticity equation and in the model by KR05 they do
not play a role in the generation of the NSSL.  

Ignoring magnetic fields, the vertical and horizontal
stresses are given by 
\EQA
Q_{r\phi}&=&\nu_{\parallel}\sin \theta \left( V\Omega - r\frac{\partial \Omega}{\partial r}\right) + \nu_{\perp}\Omega^2\sin\theta^2\cos \theta \frac{\partial
  \Omega}{\partial \theta}, \label{eq:qrp}\\
Q_{\theta\phi}&=&\nu_{\parallel}\!\left(\!\cos \theta  H \Omega\!-\!\sin \theta \frac{\partial
  \Omega}{\partial \theta} \right)\!+\!\nu_{\perp}\Omega^2\sin\theta^2\cos \theta
r\frac{\partial \Omega}{\partial r},\label{eq:qtp}
\ENA
where $\nu_{\parallel}$ and $\nuper$ are the diagonal and
off--diagonal components of the turbulent viscosity tensor $N_{ijkl}$,
respectively. The latter component $\nu_{\perp}$ appears due to the
effect of the rotation on the turbulent motions \citep{RKKS19}. $V$
and $H$ are the vertical and horizontal $\Lambda$--effect
coefficients which are, to the lowest order, proportional to $A_{\rm
  V}$ and $A_{\rm H}$ \citep{GR80}. These coefficients are typically
expanded in latitude in powers of $\sin^2 \theta$ as
\EQA
V=\sum_{i=0}^{j}V^{(i)}\sin^{2i}\theta,\label{eq:v}\\
H=\sum_{i=1}^{j}H^{(i)}\sin^{2i}\theta.\label{eq:h}
\ENA
In the NSSL $\con\leqslant 1$ and $A_{\rm H} \approx 0$ such that
$Q_{\theta\phi}^{(\Lambda)}$ due to the $\Lambda$--effect
vanishes. The off--diagonal viscosity $\nuper$ is non--zero but small
such that its influence is negligible \citep{RKKS19}.
It has been shown analytically \citep{LKGR05} and numerically
\citep{Kap19} that in the slow rotation regime only the first term in
the expansion of the vertical coefficient $V^{(0)}$ survives and tends
to a constant. Furthermore, applying a stress--free boundary condition
at the radial boundaries, one realizes that $Q_{r\phi}=Q_{r\theta}=0$.
Using this in Eq.~(\ref{eq:qrp}) and equating the diffusive and 
non--diffusive stresses we get
\EQ
\frac{\partial\ln \Omega}{\partial \ln r}=V^{(0)}<0,
\EN
which shows a reasonable agreement with observational results where
the radial rotational gradient is independent of latitude \citep{BSG14}.
\begin{figure}[t!]
\begin{center}
\includegraphics[width=0.8\columnwidth]{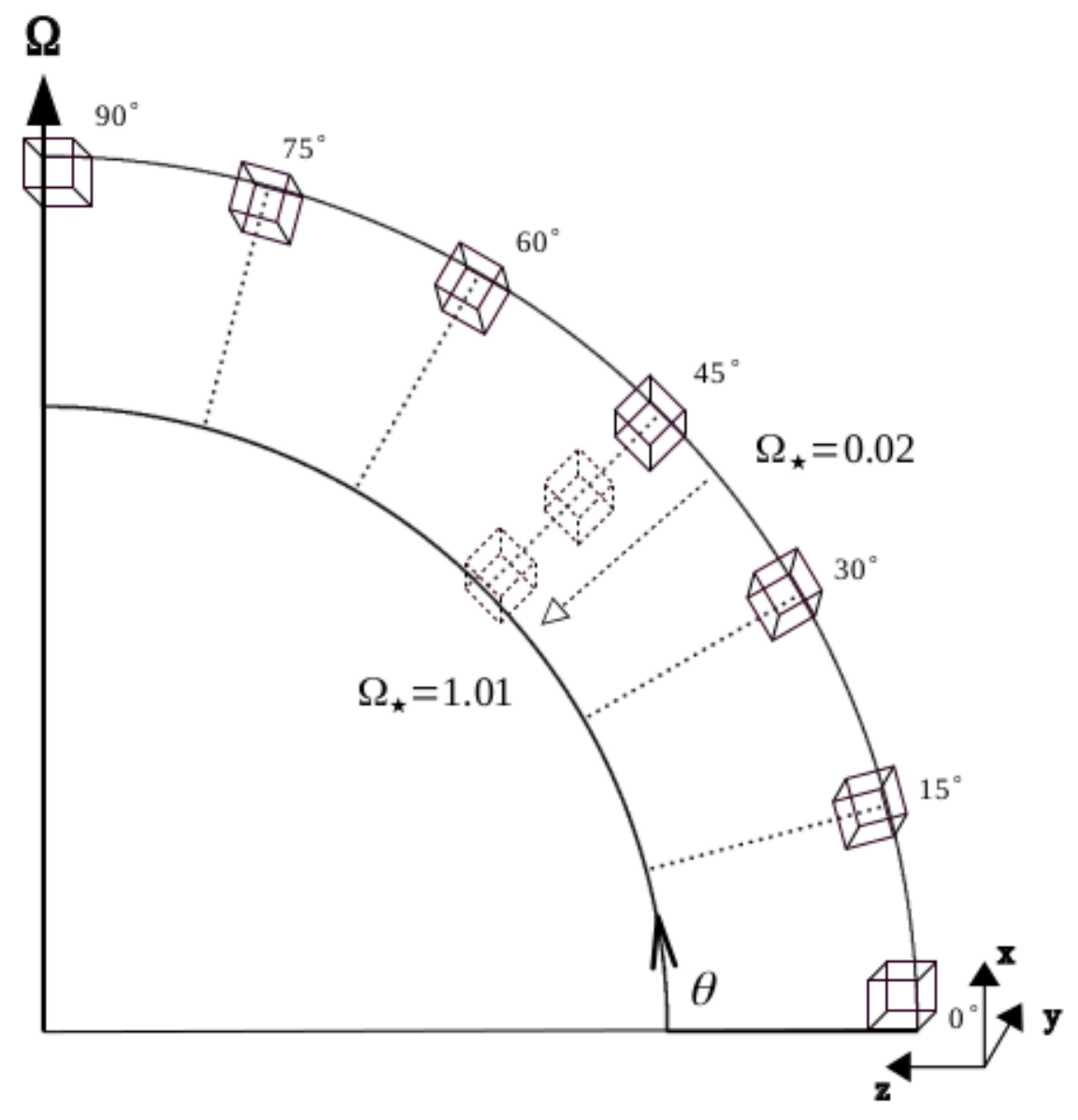}
\end{center}\caption[]{Schematic representation of geometry of the
  current models and their relation to the NSSL. The depth of the
  layer is exaggerated. The simulation boxes are located at nine
  depths (not all shown) and seven latitudes, respectively. $\con$ is
  increasing gradually from the surface to the bottom of the NSSL.
}\label{fig:sk}\end{figure}
\section{The model}
\label{sec:model}

We use a similar hydrodynamic model in Cartesian domain as in
\cite{KB08} and \cite{Kap19}. We explain it here briefly and refer the
reader to relevant parts of the above--mentioned works for details. In
this model gravity is neglected and an external random forcing by
non--helical transversal waves with direction--dependent amplitude is
applied. 
The ensuing flow is turbulent and anisotropic. The medium is
considered to be isothermal and to obey the ideal gas equation. The
governing equations are   
\EQA
\frac{D \ln \rho}{Dt}&=&-\nab \bm\cdot \UU,\\
\frac{D\UU}{Dt}&=&-c_s^2 \nab \ln \rho-2\ \bm\Omega\times\UU  +\bm{F}^{\rm
  visc} +\bm{F}^{\rm f},
\label{momentum}
\ENA
where $D/Dt=\partial / \partial t+\UU \bm\cdot \nab$ is the
advective derivative, $\rho$ and $c_s$ are the density and sound
speed, respectively, and
$\bm\Omega=\Omega_0(-\cos\theta,0,\sin\theta)^{\rm T}$ is the rotation
vector. The viscous force is given by
\EQ
\bm{F}^{\rm visc}=\nu\left(\nabla^2\UU+\frac{1}{3} \nab \nab \bm\cdot
\UU+2\SSS \bm\cdot \nab \ln \rho \right),
\EN
where $\SSS_{ij}=\frac{1}{2}( U_{i,j}+ U_{j,i})-
\frac{1}{3}\delta_{ij}U_{k,k}$ is the traceless rate of strain
tensor, $\delta_{ij}$ is the Kronecker delta, and the commas denote
differentiation. The forcing function is given by
\EQ
\bm{F}^{\rm f}( \boldsymbol{x},t)={\rm Re} (\NNNN \cdot
\boldsymbol{f_{k(t)}} \exp [i \boldsymbol{k}(t)\bm\cdot \boldsymbol{x}
  - i\phi(t)]), 
\EN
where $ \boldsymbol{x}$, $\boldsymbol{k}$, and $\phi$ are the position,
wavevector, and a random phase, respectively. The desired vertical ($z$)
anisotropy can be enforced using a tensorial normalization factor 
${\sf N}_{ij}=(f_0\delta_{ij}+\delta_{iz} \cos^2 \Theta_k f_1/f_0) 
(kc_s^3/\delta t)^{1/2}$ of the forcing, where $f_0$ and $f_1$ are the
amplitudes of the isotropic and anisotropic parts, respectively. $\delta
t$ and $\Theta$ are the time step and the angle between the vertical
direction $z$ and $\boldsymbol{k}$, respectively, and $k=|\boldsymbol{k}|$ 
determines the dominant size of the eddies. In the forcing
$\boldsymbol{f_{k}}$ is given by
\EQ
\boldsymbol{f_{k}}=\frac{\boldsymbol{k}\times \boldsymbol{\hat{e}}
}{\sqrt{\boldsymbol{k}^2-(\boldsymbol{k\cdot\hat{e}})^2}},
\EN
which makes the forcing transversal waves; $\hat{\bm e}$ is an arbitrary
unit vector. The details of the forcing can be found in
\cite{AB01}. 
\begin{table}[t!]\caption{ 
Summary of runs of varying the Taylor number and latitude.
The values of $A_V$ are shown at the equator and the pole and the
values of $A_H$ at $15^\circ$ and $75^\circ$ latitude.  
}\vspace{12pt}\centerline{\begin{tabular}{lcccc}                            
\hline          
set & Ta $(10^6)$ & $\con$ & $A_{\rm V}$ $(0^\circ ... 90^\circ)$ &
$A_{\rm H} [10^{-3}]\,(15^\circ ... 75^\circ)$\\ 
 \hline\hline
 \textbf{C0} & 0 & 0 & -0.51 & 0.56\\
 \textbf{C02} & 0.05 & 0.02 & -0.51 & $0.73 ... 0.57$\\
 \textbf{C04} & 0.15 & 0.04 & -0.51 & $0.72 ... 0.66$\\
 \textbf{C06} & 0.30 & 0.06 & -0.51 & $0.16 ... 0.32$\\
 \textbf{C13} & 1.40 & 0.13 & -0.51 & $0.73 ... 0.30$\\
 \textbf{C24} & 4.54 & 0.24 & -0.51 & $2.85 ... 0.19$\\
 \textbf{C46} & 15.58 & 0.46 & $-0.52 ... -0.50$ & $8.84 ... 1.41$  \\
 \textbf{C64} & 30.54 & 0.64 & $-0.52 ... -0.49$ & $14.3 ... 1.12$ \\ 
 \textbf{C83} & 50.49 & 0.83 & $-0.52 ... -0.48$ &  $20.7 ... 1.11$ \\
 \textbf{C1} & 75.43 & 1.01 & $-0.52 ... -0.46$ &  $27.0 ... 0.89$ \\
\hline
\label{tab:summary}\end{tabular}}
\tablefoot{
The grid resolution of all runs is $144^3$, forcing amplitudes
$f_0=10^{-6}$ and $f_1=0.04$, ${\rm Re}\approx 13$, and
$\nu=3.3\cdot10^{-4}~(c_s k_1^3)^{-1}$.
}
\end{table}
\section{Simulation setup}
\label{sec:setup}

We used the \PC \footnote{https://github.com/pencil-code} 
\citep{PC20} to run
the simulations. We consider a cubic box with size $(2\pi)^3$ 
discretized over 144$^3$ grid points. $z$ corresponds to vertical, 
$x$ to latitudinal, and $y$ to azimuthal direction,
respectively, the latter two being referred to as the
horizontal directions. Horizontal boundaries are periodic
and stress--free conditions are imposed at vertical boundaries with   
\EQ
U_{x,z}=U_{y,z}=U_z=0 \quad\mbox{on   $\quad z=z_{\rm bot}$, $z_{\rm top}$},
\EN
where $z_{\rm bot}$ and $z_{\rm top}$ represent the bottom and top of the domain.
The box size is represented by the wavenumber $k_1=2\pi/L$ and we choose a 
forcing wavenumber $k_f/k_1=10$. The units of length, time, and
density are $k_1^{-1}$, $(c_sk_1)^{-1}$ and $\rho_0$, 
respectively, where $\rho_0$ is the initial uniform value of density.
The forcing parameters $f_0=10^{-6}$ and $f_1=0.04$ are chosen such that 
the effects of compressibility are weak with a Mach number ${\rm
  Ma}=\urms/c_s \approx 0.04$ in all simulations. Moreover, with $f_1 \gg f_0$, 
we fulfill the NSSL condition in which $|A_{\rm V}|\gg |A_{\rm H}|$;
see \Tab{tab:summary}. 
The vigor of turbulence is quantified by the Reynolds number
\EQ
{\rm Re}=\frac{\urms}{\nu k_f},
\EN
where $\urms=(\overline{\UU^2}-\overline{\UU}^2)^{1/2}$, is the root 
mean square of the fluctuating velocity field. Using a fixed value of
the kinematic viscosity, $\nu=3.3\cdot10^{-4}~(c_s k_1^3)^{-1}$, the  
Reynolds number is about 13 for all simulations.
We place the box at seven equidistant latitudes from the equator to
the pole by setting the angle $\theta$ between the rotation vector
and the vertical direction as shown in \Fig{fig:sk}. The vertical 
placement is determined by the value of $\Omega_0$ which is varied 
such that the range of $\con$ from \Eq{eq:con} is relevant for the 
NSSL. The turnover time is defined as $\tau=\ell/\urms$, where $\ell$
is the size of the eddies. In our simulations the energy--carrying
scale of turbulence is the forcing scale $\ell=2\pi/k_f$. Hence, the
Coriolis number in the simulations is given by
\EQ
\con=\frac{4\pi\Omega_0}{\urms k_f}.
\EN
The corresponding input parameter is the Taylor number
\EQ
{\rm{Ta}}=\left (\frac{2\Omega_0L^2}{\nu}\right)^2.
\EN
The values of ${\rm Ta}$, $\con$, and the anisotropy parameters are
given in \Tab{tab:summary}. An additional run with $\Omega_0=0$ was 
performed to remove a contribution to the Reynolds stress appearing
in the non--rotating case; see \Sec{res:RS}.

Mean quantities are defined as horizontal ($xy$) averages. 
The local Cartesian quantities are related to their counterparts in 
spherical polar coordinates via
$(r,\theta,\phi)\rightarrow (z,x,y)$,
$(\meanUU_r,\meanUU_{\theta},\meanUU_{\phi})\rightarrow
(\muz,\mux,\muy)$, $Q_{\theta \phi}\rightarrow \qxy $,  
$Q_{\theta r}\rightarrow \qxz$ and $Q_{r \phi }\rightarrow \qyz  $.
We normalize quantities such that
$\widetilde{\UU}_{i}=\meanUU_i/\urms$ and
$\widetilde{Q}_{ij}=Q_{ij}/\urms^2$, tilde denoting this
operation. Additionally, the error on the measured physical
quantities, which are obtained directly from the simulations, is
estimated by dividing the time series into three parts and comparing
their time averaged values with the one obtained from the whole time
series. The maximum deviation from the latter is considered to be the
error of the measurement.  

\begin{figure}[t!]
\begin{center}
\includegraphics[width=\columnwidth]{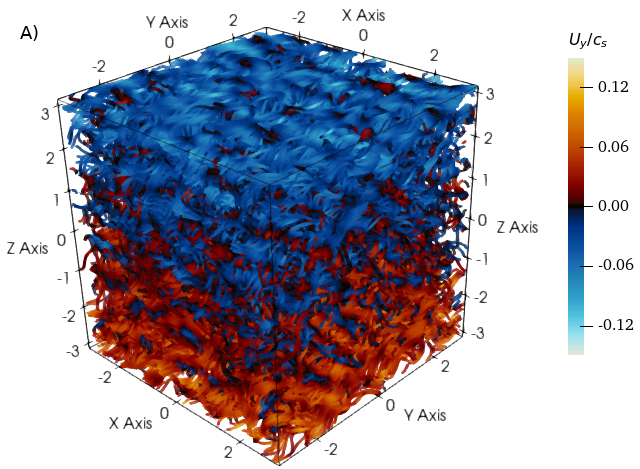}
\includegraphics[width=\columnwidth]{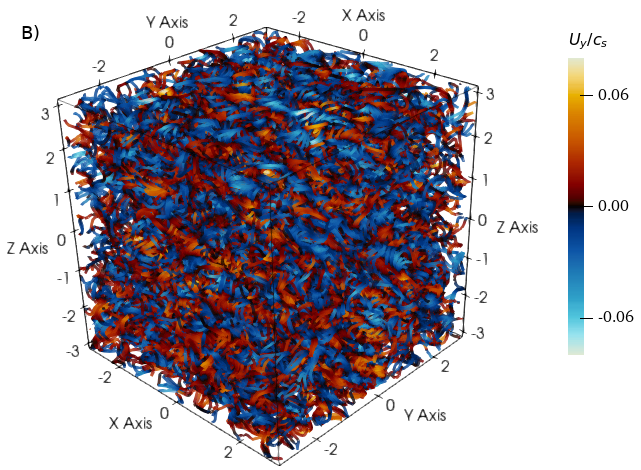}
\end{center}\caption[]{Streamlines of the velocity field. The color
  table shows the amplitude of the azimuthal component of the velocity
  field normalized by sound speed. Panels A and B show $U_y/c_s$ at
  the equator and at $\theta = 30^\circ$ for set \textbf{C46},
  respectively.  
}\label{fig:Uy}\end{figure}

\begin{figure}[t!]
\begin{center}
\includegraphics[width=\columnwidth]{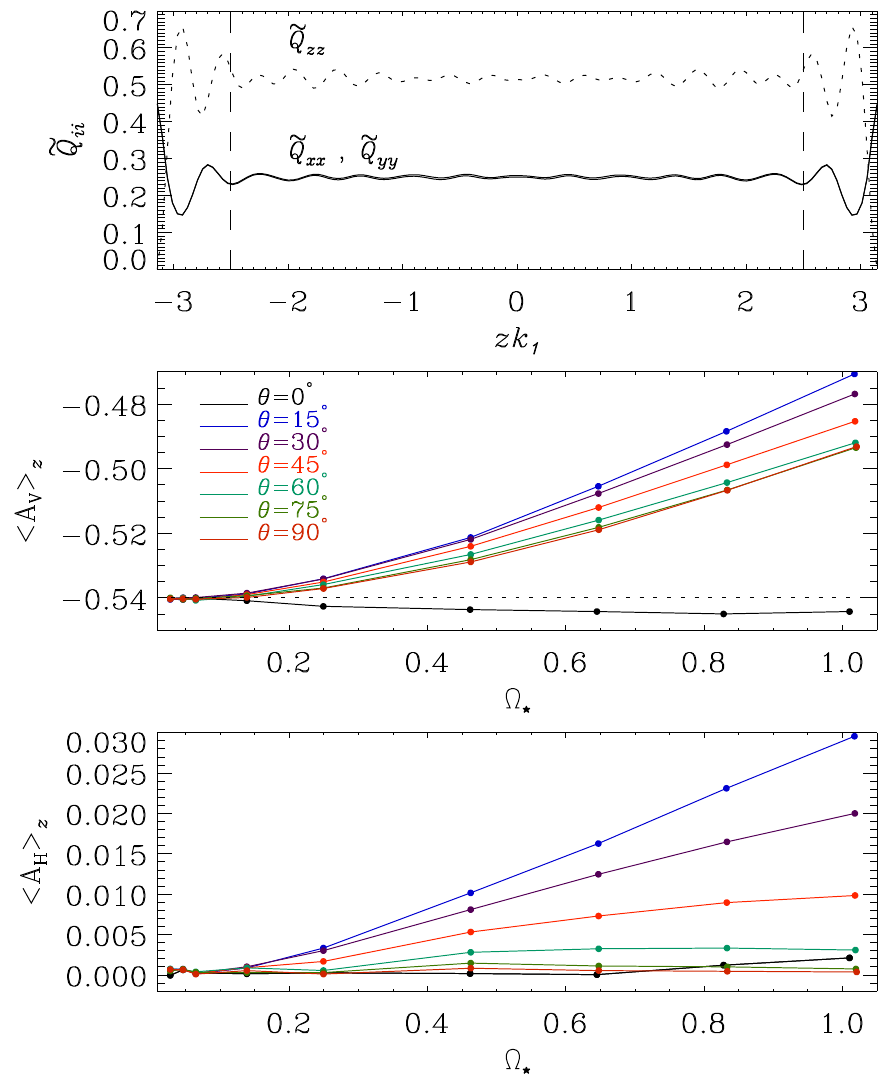}
\end{center}\caption[]{Top panel: Time averaged and normalized diagonal 
components of the Reynolds stress as functions of $z$.
  The dotted (solid) line shows
  $\tqzz$ ($\tqxx$ and $\tqyy$) of the set
  \textbf{C24} at $15^\circ$ latitude. The vertical dashed lines mark
  the part of the domain wherefrom $A_{\rm  V}$ and $A_{\rm H}$ were measured. 
  Anisotropy parameters $A_{\rm V}$ (middle panel) and $A_{\rm H}$
  (bottom panel) are shown as functions of $\con$ at the latitudes
  indicated in the legend. 
}\label{fig:AvAh}\end{figure}
\section{Results}
\subsection{Velocity field}

A statistically stationary turbulent state appears after about few $\tau$ 
independent of $\con$ everywhere except at the equator, where  
the statistically stationary state is reached between few to about 300 
$\tau$ from the lowest to highest $\Omega_\star$, respectively.
As an example, we show snapshots of the zonal flow
normalized by the sound speed at about 1000~$\tau$ for the set 
\textbf{C46} at the equator and at $30^\circ$ latitude in panel A and
B of \Fig{fig:Uy}, respectively. 
The other components of the velocity field are very similar to
the zonal one shown in panel B. The dominant scale of the turbulence is 
the forcing scale $k_f/k_1=10$.
The expected large--scale zonal flow similar to the actual NSSL is
generated only at the equator shown in panel A. All other sets show
similar behaviour.  

\begin{figure*}[t!]
  \begin{center}
\includegraphics[width=0.75\textwidth]{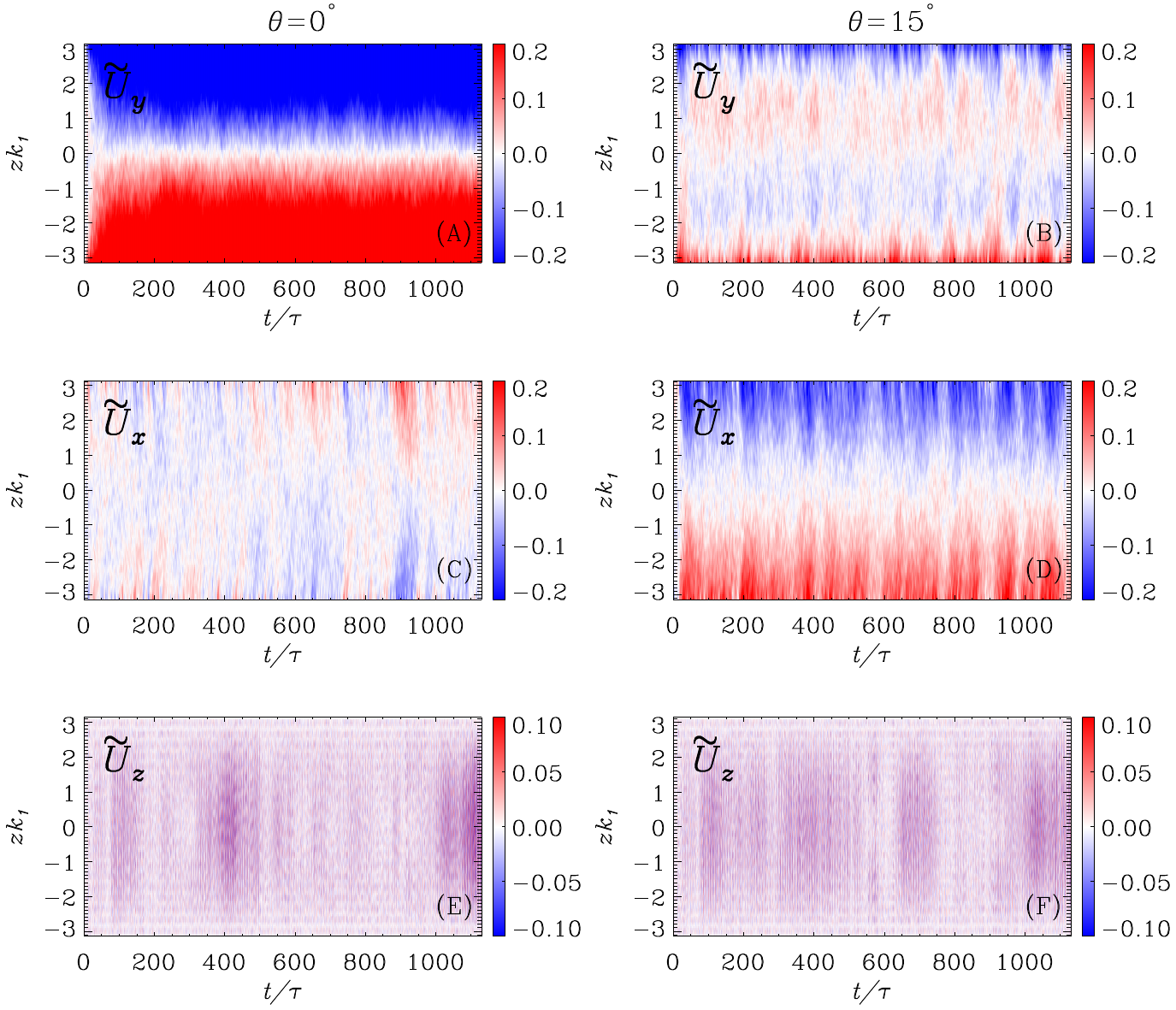}
\end{center}\caption[]{Normalized mean components of the velocity
  field versus time in terms of turnover time in representative runs in 
  set \textbf{C24}. The rows from top to bottom show $\tmuy$,  $\tmux$ and
  $\tmuz$, respectively. The left and right columns show the
  mean velocities at the equator and at $15^\circ$ latitude, respectively. 
  To make the comparison of the velocity components feasible, we clip
  the values of the color table of panel (A) at 50 per cent of the
  maximum value. 
}\label{fig:uxy27}\end{figure*}
\subsection{Anisotropy of the flow}
\label{sec:Aniso}
We start our analysis by measuring the diagonal components of the
Reynolds stresses and the anisotropy parameters which are given by    
\EQA
A_{\rm V}&=&\frac{Q_{xx}+Q_{yy}-2Q_{zz}}{\urms^2},\\
A_{\rm H}&=&\frac{Q_{yy}-Q_{xx}}{\urms^2}. 
\ENA
We show representative time--averaged diagonal stresses in the top
panel of \Fig{fig:AvAh}. The stresses are almost constant in the entire
domain except at the boundaries, where $\qzz=0$ and the horizontal
components rise to twice larger values. Additionally, in the interior
the values of $\tqzz$ are about twice as large as the other two
components, reflecting the fact that $A_{\rm V}\approx -0.5$.

We show the volume averaged $A_{\rm V}$ and $A_{\rm H}$ as a function
of $\con$ at different latitudes in the middle and bottom panels of
\Fig{fig:AvAh}, respectively. 
We consider data between $-2 \le z k_1 \le 2$ for the volume averages
to avoid boundary effects.
The vertical anisotropy parameter $A_{\rm V}$ is always at least 
two orders of magnitude greater than $A_{\rm H}$. Neither shows 
appreciable variation as a function of Coriolis number for 
$\Omega_\star \lesssim 0.15$.
The vertical anisotropy parameter is almost independent of $\con$ at
the equator in contrast to other latitudes where its absolute value
decreases with increasing $\con$. It decreases
about 15\% at the bottom of the NSSL at $15^\circ$ and about 5\% at
latitudes above $45^\circ$.
The horizontal anisotropy parameter shows almost no dependence on
latitude above $45^\circ$ but it becomes 100 times greater
from the top to the bottom of the NSSL below this latitude.
The behaviour of both anisotropy parameters is similar to the ones
obtained by \cite{KB08} in which they have similar set--up as ours but
applied fully periodic boundary conditions. This shows that anisotropy
of the flow is insensitive to the boundary conditions.

\subsection{Mean flows}
\begin{figure}[t!]
\begin{center}
\includegraphics[width=\columnwidth]{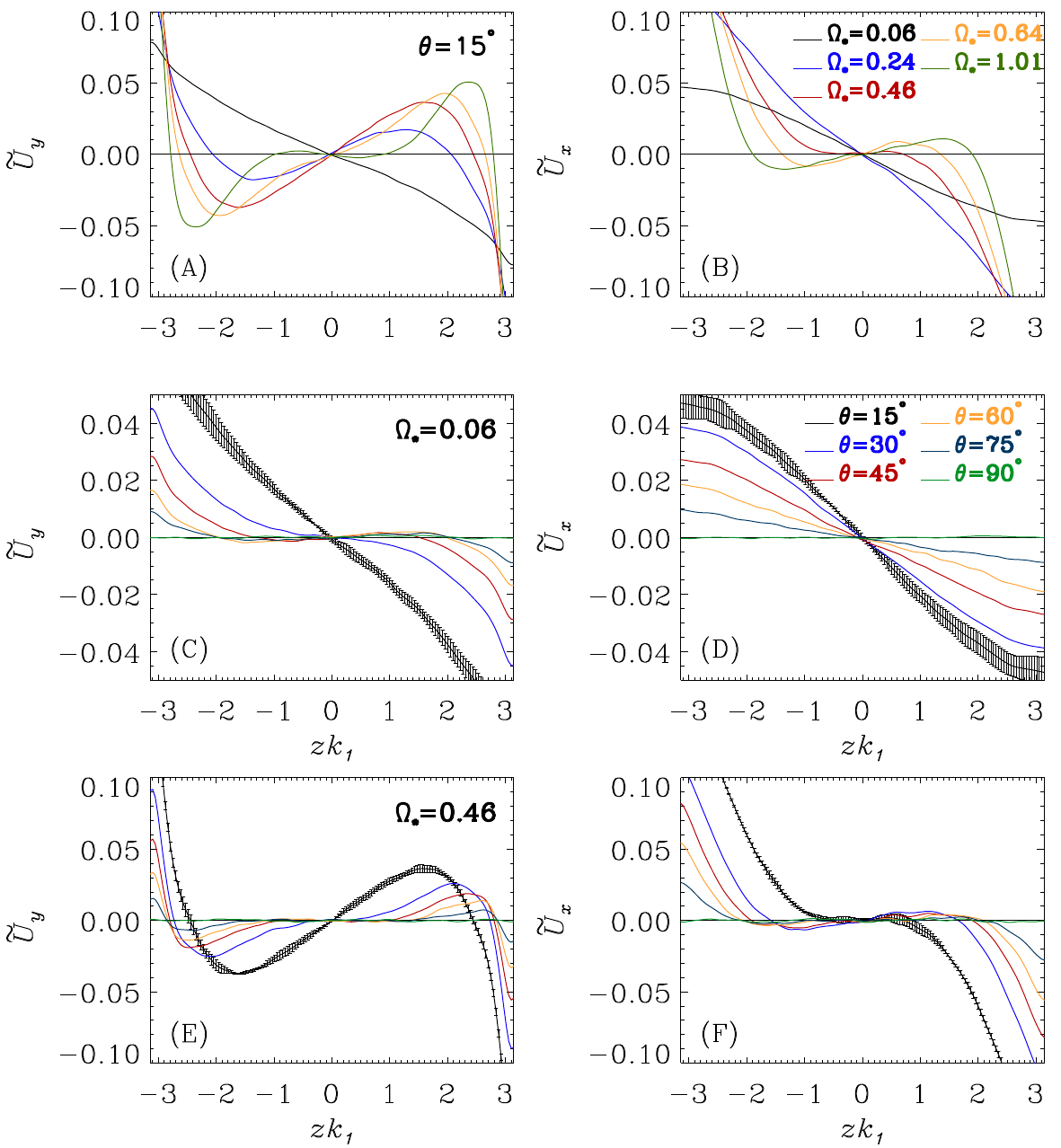}
\end{center}\caption[]{ Time--averaged normalized mean
  velocity components versus vertical direction. The panels (A) and
  (B) show $\tmuy$ and $\tmux$ at $15^\circ$.
  The second and third rows show the mean horizontal velocities for sets
  \textbf{C06} and \textbf{C46}, respectively, at the latitudes indicated 
  by the legends.  
}\label{fig:muco}\end{figure}

The development of mean flows in rotating cases means that reaching a 
statistically steady state takes significantly longer than in non--rotating 
runs. Furthermore, long time averages are needed for statistical convergence 
of the turbulent quantities.
We run all the simulations for at least 1100 turnover times. As an
example, we show a subset of the time evolution of the three components of the
normalized mean velocity field for about 1200 turnover times 
for the set \textbf{C24} at the equator and at $\theta=15^\circ$ in
\Fig{fig:uxy27}. 
At the equator, a large zonal flow $\muy$ with a negative vertical
gradient developed gradually over $100\tau$ as shown in panel (A). 
All other sets show similar zonal flow profile at the 
equator, but both the amplitude and steepness of the gradient
increase with increasing $\con$.
Moving away from the equator, the amplitude of the mean zonal flow
reduces significantly and the negative gradient disappears
as shown in panel (B) of \Fig{fig:uxy27}. The dependence of mean zonal
flow on rotation can be seen in the panel (A) of \Fig{fig:muco} where
we show the time--averaged $\tmuy$ at selected $\con$ at $15^\circ$
latitude. By increasing $\con$, the gradient of $\tmuy$ changes sign
and becomes steeper up to $\con=0.46$, then it becomes shallower and
slowly vanishes in the middle at $\con=1$. 
The latitudinal dependence of $\tmuy$ is shown for
sets \textbf{C06} and \textbf{C46} in the panels (C) and (E) of
\Fig{fig:muco}, respectively. We find that $\tmuy$ decreases as a
function of latitude, vanishes at the poles, and that the amplitude is
less than 5\% of $\urms$ everywhere apart from the boundaries.

\label{result}
The time--averaged meridional component of the mean flow $\mux$ is
consistent with zero at the equator for all runs similar to the one is
shown in panel (C) of \Fig{fig:uxy27}. In contrast to the zonal flow,
its value increases by moving away from the equator; see panel (D) of
\Fig{fig:uxy27}. The time--averaged value of this component at
$15^\circ$ is shown in panel (B) of \Fig{fig:muco} for selected values
of $\con$. The negative gradient persists up to $\con=0.24$. Above
this $\con$, the shear slowly vanishes at the center of the box and
becomes slightly positive by increasing $\con$. However, the strong
shear persists only near the boundaries. We show the latitudinal
dependency of $\mux$ for the two sets \textbf{C06} and  \textbf{C46}
in panels (D) and (F) of \Fig{fig:muco}. The amplitude of $\mux$
decreases as a function of latitude. The amplitudes of $\mux$ and
$\muy$ are comparable everywhere apart from the equator and the
negative gradient of $\mux$ for $\con<0.1$ persists at all latitudes.

For completeness, we show the vertical component of the normalized mean 
flow $\tmuz$ in the bottom row of \Fig{fig:uxy27}. All runs show a similar 
pattern of high--frequency oscillations for $\tmuz$ irrespective of
latitude and $\con$ with amplitudes of the order of
$10^{-4}\urms$. These oscillations are identified as longitudinal 
sound waves as expected for a compressible system in a confined cavity.

\subsection{Reynolds stresses}
\label{res:RS}
\begin{figure}[t!]
\begin{center}
\includegraphics[width=\columnwidth]{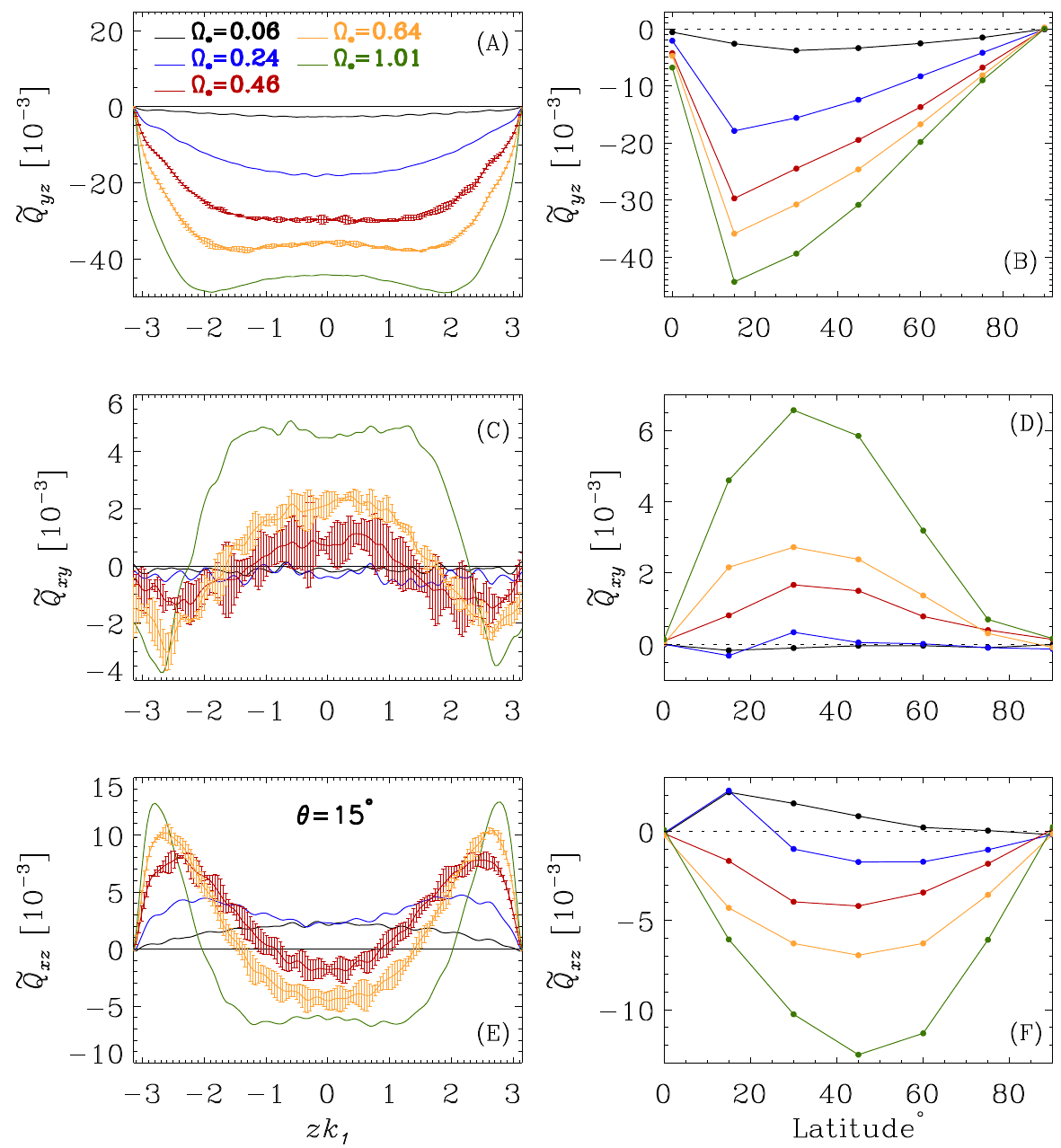}
\end{center}\caption[]{ 
Left column: Time--averaged off--diagonal Reynolds stresses versus
vertical direction at 5 selected $\con$ indicated by the
legends. Right column: The stresses shown on left panels further
spatially averaged ($-0.5 \le z k_1 \le 0.5$), at different
latitudes. The rows from top to bottom show $\tqyz$,  $\tqxy$ and
$\tqxz$, respectively. 
}\label{fig:allq}\end{figure}

For zero rotation, it is expected that $\tqxy = 0$, see \Eq{eq:qtp}. However, we
find that $\tqxy$ always has a small but non--zero value which persists also in
the longest time series of our data. We find that such contribution is present
and its magnitude remains unchanged also for higher resolutions
($288^3$ and $576^3$ grids).  
Hence, this issue seems not to be caused by a numerical convergence
issue, but we have been unable to identify whether the cause is due to
some, yet unidentified, physical effect for example due to
compressibility, effects due to the forcing, inhomogeneities in the
system, or a combination thereof. Given that this contribution is
systematically present, we made a non--rotating run (\textbf{C0}) and
subtracted $\tqxy$ from that run from the results of the runs with
rotation. 

We show representative results of the off--diagonal stresses at five selected
$\con$ at $15^\circ$ (left column) and as function of latitude (right
column) in \Fig{fig:allq}. The data for all sets is available as
online electronic material.
The vertical Reynolds stress at all latitudes shows similar profiles as at 
$15^\circ$, see panel (A). The stress is nearly constant in the
interior of the domain and tends to zero at the boundaries. 
$\tqyz$ is always negative independent of $\con$ and latitude, as
shown in panel (B). Thus, the vertical angular momentum transport is inward
in agreement with previous studies
\citep[e.g.][]{PTBNS93,Chan2001,KKT04,LKGR05,KB08,Kap19}. Independent
of $\con$, the vertical stress vanishes at the pole and has its
minimum and maximum amplitude at the equator and $15^\circ$,
respectively, after which it decreases gradually towards the pole. 
For a given $\con$, its amplitude is about twice larger at $30^\circ$ latitude 
than $60^\circ$. The latitudinal dependence of $\qyz$ is different
from previous studies by \cite{PTBNS93} and \cite{KKT04} at
$\con\approx 1$ in which they measured $\qyz$ from local box
convection simulations. In \cite{PTBNS93} it is almost constant up to
60$^\circ$ and decreases toward higher latitudes. $\qyz$ has a
v--shape profile in latitude with the minimum at $45^\circ$ in
\cite{KKT04}. The major ingredient which is missing in our forced
turbulence simulation in comparison with theirs is density
stratification. Moreover, \cite{KKT04} includes the overshooting layer
below the CZ. Therefore, it is difficult to find out what makes our
results different from theirs.

The middle panels (C) and (D) in \Fig{fig:allq} show horizontal stress
$\qxy$. The signature of turbulent fluctuations at the forcing scale
are seen more clearly in this component and the measurement is quite
noisy. The values of $\qxy$ are close to zero up to $\con=0.46$, above
which it slowly starts to get positive (negative) values in the middle
(close to the boundaries). This is the same behavior as with $A_{\rm
  H}$ seen in \Fig{fig:AvAh}. At a given $\con$, the profile of $\qxy$
is similar at all latitudes. Its amplitude is maximum at $30^\circ$ and decreases
gradually towards the equator and the pole as shown in panel (D). This
result is in agreement with the observational measurements of $\qxy$
using sunspot proper motions \citep{FW65,GH84,PT98}, but not with the
one measured using supergranulation motions, see Figure 10 in
\cite{HGS16}. The horizontal stress has always positive values
independent of $\con$ and latitude in agreement with previous studies
in slow rotation regime \citep[e.g.][]{LKGR05,KB08,Kap19}. The
latitudinal profile of $\qxy$ measured by \cite{PTBNS93} is very
similar to our results, albeit with negative values as their box is
located at the southern hemisphere; see their Figure 6.

The meridional stress is shown in the last row of \Fig{fig:allq}. 
In contrast to the other stresses, $\tqxz$ shows a complicated
profile, in particular close to the boundaries. Moreover, it has
positive or negative values depending on both $\con$ and $\theta$
shown in panel (E). The latitudinal dependency of the meridional
stress is shown in panel (F). At $\con<0.1$, $\tqxz$ is positive at
low latitudes and $\tqxz \rightarrow 0$ above $45^\circ$. 
By increasing $\con$, $\tqxz$ moves toward negative values and its
absolute value increases. For $\con>0.24$, it has its maximum
amplitude at about $45^\circ$ and it decreases toward the pole and the
equator similar to $\tqxy$.   
The meridional stress in \cite{PTBNS93} also shows a sign change in
agreement with ours albeit in mid latitude. However, the sign change
occurs at $\con\approx 1$ while our shows only negative values at that
$\con$.   

Comparing the absolute amplitude of the stresses in right column of
\Fig{fig:allq} we see that $\tqyz$ always larger than $\tqxz$ and
$\tqxy$. For example at $\con=0.64$, $\tqyz$ is about two to ten times
larger than $\tqxz$ and five to twenty times larger than $\tqyz$
depending on latitudes. Comparing also the absolute amplitude of  
$\tqxy$ and $\tqxz$, we see that $\tqxz>\tqxy$ for all $\con$.
 These results show that, in spite of the fact that $\qxy$ is
 increasing as a function of $\con$, its values are still much smaller
 than vertical stresses which is in agreement with the assumption of
 KR05 regarding the NSSL. 

Although our model is quite simple in comparison to GDNS, it is of
interest to compare the Reynolds stresses with simulations such as
those in \cite{KMGB11}.  
These authors modelled turbulent convection in a spherical wedge for a
variety of rotation rates. Considering the runs of \cite{KMGB11} with
$\con<1$ we find good agreement for the horizontal stress $\qxy$ which
is small and positive for small $\con$, and which has appreciable
values only for $\con>0.5$. However, we find maximal values at
$30^\circ$ instead of at $10\ldots15^\circ$ in \cite{KMGB11}. We also
observe a similar trend for $\qxz$ such that it is positive for small
$\con$ on the northern hemisphere with a sign change after certain
$\con$. However, this trend depends on latitude in their case; see
their Figure~8. The profile of $\qyz$ in the convection simulations is
quite different from ours such that it has a strong latitudinal
dependency and gets both positive and negative values depending on
$\con$ and latitude. 
This is consistent with earlier studies \citep[e.g.][]{Kap19} where a
sign change of $\qyz$ occurs at higher $\con$ than those considered in
the present simulations. 

\subsection{The role of Reynolds stresses in the generation of the mean
  flows}
\begin{figure}[t!]
\begin{center}
\includegraphics[width=\columnwidth]{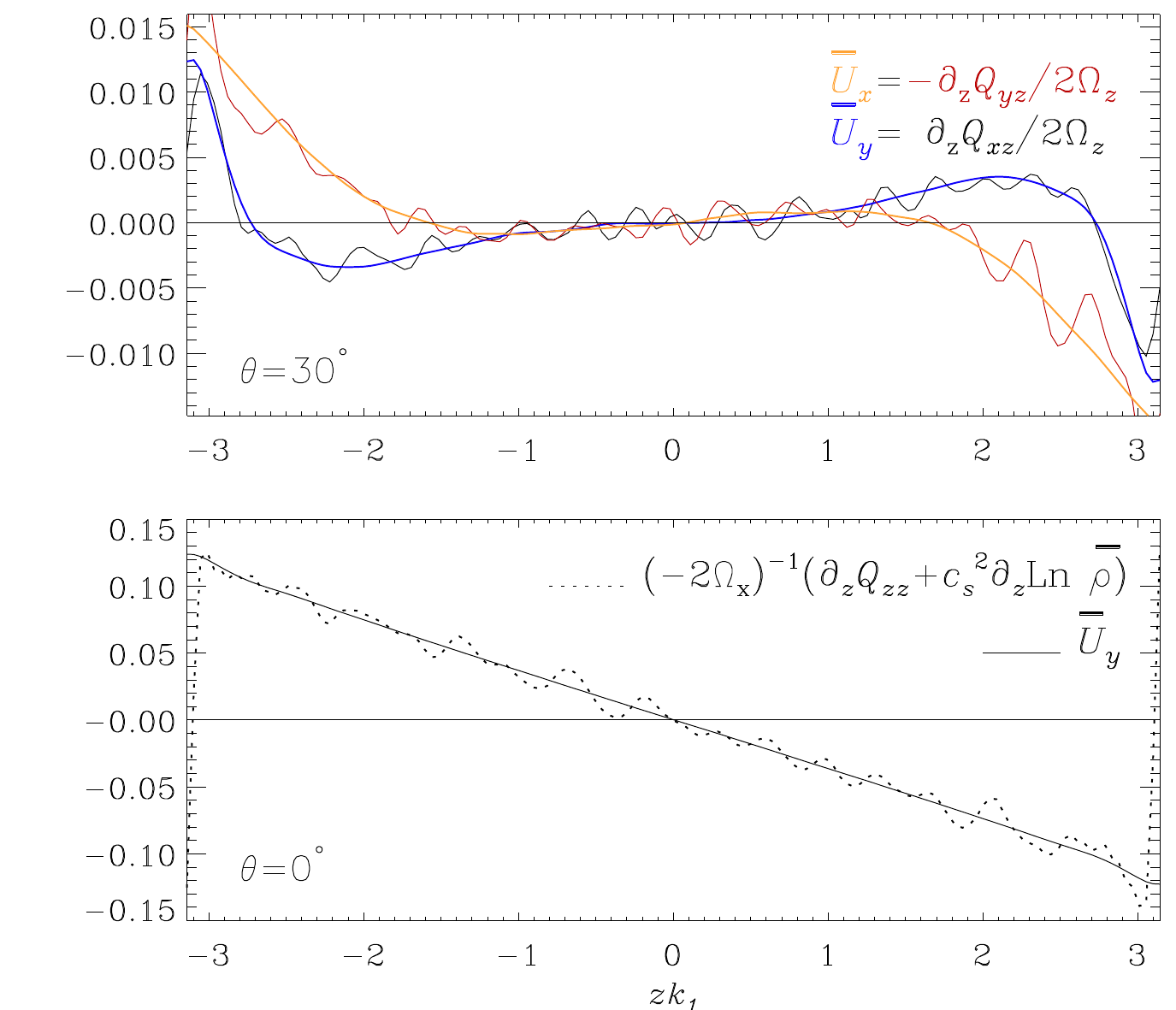}
\end{center}\caption[]{The mean velocities $\muy$ and $\mux$ and their
  corresponding balancing terms in \Eqs{eq:bmuy}{eq:bmux} at $30^\circ$
  latitude (upper panel) and
  \Eq{eq:bbmuy} at the equator (lower panel) over vertical direction for
  set \textbf{C46}. In the upper  panel, the orange and blue lines show
  $\mux$ and $\muy$, 
  respectively. The red and black lines show the RHS of
  \Eqs{eq:bmux}{eq:bmuy}, respectively. In the lower panel, the solid and
  dotted lines show the LHS and RHS of \Eq{eq:bbmuy}, respectively.  
}\label{fig:balance}\end{figure}

As the Reynolds stresses appear in the 
MF momentum equation, we start by writing the MF equations for
$\mux$ and $\muy$ using \Eq{momentum}.
We wrote these equations first by considering the facts that our setup
is fully compressible and the forcing we used here is not solenoidal
which might cause  density fluctuations which cannot be ignored in the
MF equations. These considerations lead to the presence of three
additional terms to the Reynolds stresses in the momentum equation
\citep[e.g.][]{KRBK20}. We compared all 
of them with the Reynolds stresses, and it turns out that they, and
their gradients, are considerably smaller than the Reynolds stresses.
Therefore, we can ignore the density fluctuations, and the final set of 
equations reads
\begin{equation}
\dot{\mux}\!=\!-\muz\partial_z\mux\!-\!\partial_z
\qxz\!-\!\nu\partial_z^2\mux\!-\!2(\Omega_y\muz\!-\!\Omega_z\muy),
\label{eq:muxt}
\end{equation}
\begin{equation}
\dot{\muy}\!=\!-\muz\partial_z\muy\!-\!\partial_z
\qyz\!-\!\nu\partial_z^2\muy\!-\!2(\Omega_z\mux\!-\!\Omega_x\muz).
\label{eq:muyt}
\end{equation}
Omitting terms proportional to the small quantities $\nu$ and $\muz$, 
and $\Omega_y=0$, yields the final form of the equations:
\EQA
\dot{\mux}&=&-\partial_z\qxz +2\Omega_z\muy,
\label{eq:muxtf}\\
\dot{\muy}&=&-\partial_z\qyz-2\Omega_z\mux.
\label{eq:muytf}
\ENA
We double--checked the validity of the MF equations by considering
the steady--state solution which reads
\EQA
\muy&=&(2\Omega_z)^{-1}\partial_z \qxz \label{eq:bmuy},\\
\mux&=&-(2\Omega_z)^{-1}\partial_z \qyz.\label{eq:bmux}
\ENA
We show the horizontal mean velocities in comparison with the RHS of
Eqs.~(\ref{eq:bmuy}) and (\ref{eq:bmux}) from 30$^\circ$ in set  
\textbf{C46} in the upper panel of \Fig{fig:balance}. These results are 
representative of all non--equatorial cases. Although there are 
fluctuations in the gradient of the Reynolds
stresses, the match is satisfactory.

\begin{figure}[t!]
\begin{center}
\includegraphics[width=\columnwidth]{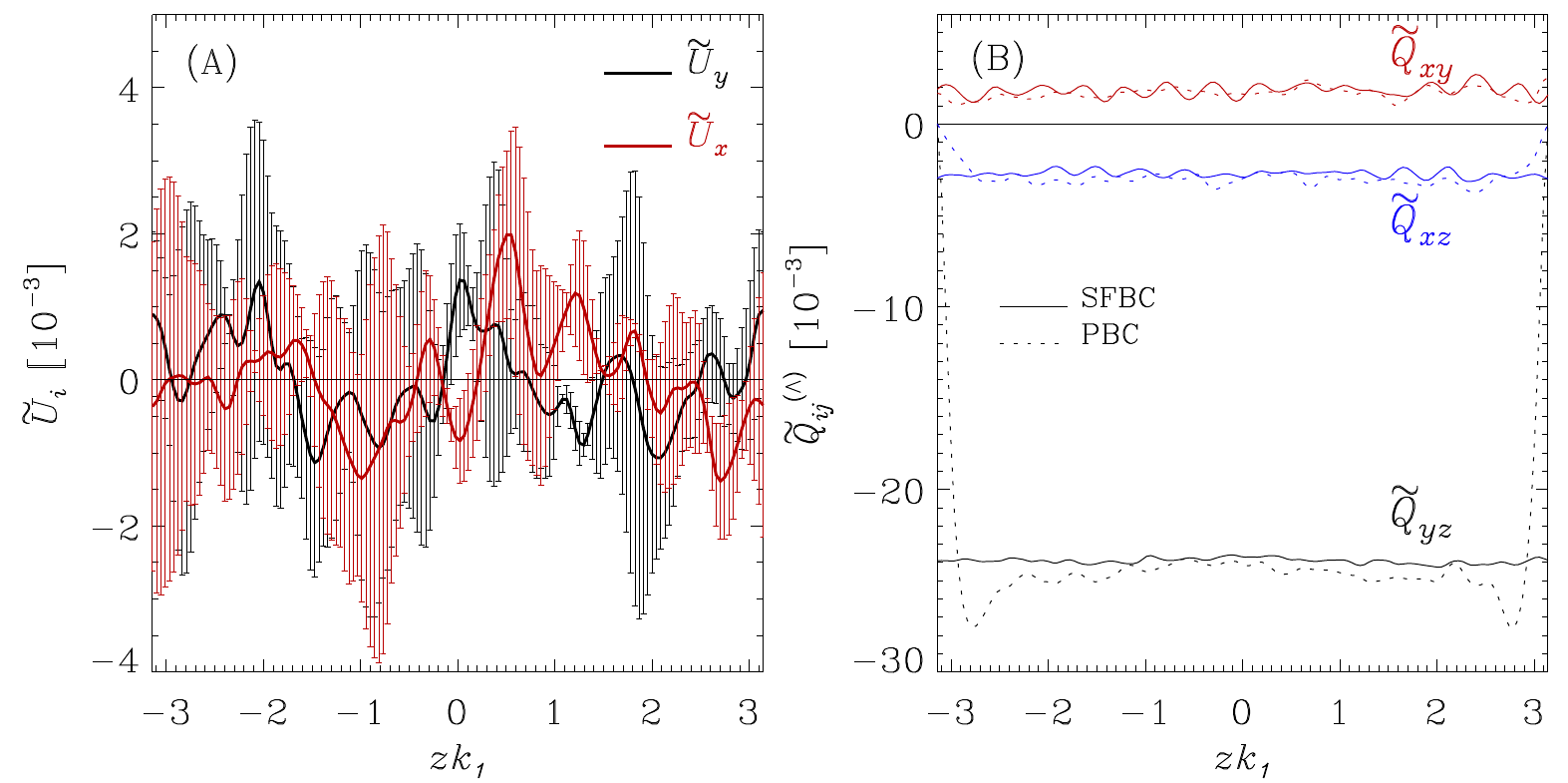}
\end{center}\caption[]{
  Panel (A): time--averaged normalized mean velocity versus vertical
  direction of periodic boundary condition (PBC) run for set
  \textbf{C46} at 30$^{\circ}$ latitude. The black and red 
  lines show $\muy$ and $\mux$, respectively. Panel (B): comparison of
  the time--averaged normalized stresses obtained from PBC and
  stress--free boundary condition (SFBC) of the same run. The solid
  and dashed lines  show the measured  $\tqxy$ (red), $\tqxz$ (blue)
  and $\tqyz$ (black) from SFBC and PBC runs, respectively.     
}\label{fig:pbc}\end{figure}
The equator is a special case and \Eq{eq:bmuy} cannot be used because
$\qxz$ and $\Omega_z$ are both zero there.
Therefore, we need to use the third component of the MF 
momentum equation. Applying similar elimination of the terms as done for
\Eqs{eq:muxt}{eq:muyt}, we have 
\EQ
\dot{\muz}=-c_s^2\partial_z \ln \meanrho-\partial_z
\qzz-2\Omega_x\muy.\label{eq:muzt}
\EN
The pressure gradient appears in this equation due to horizontal
averaging. In the steady state, the zonal flow can be written as
\EQ
\muy=-(2\Omega_x)^{-1}(\partial_z\qzz+c_s^2\partial_z\ln \meanrho).
\label{eq:bbmuy}
\EN
We show both sides of \Eq{eq:bbmuy} in the lower panel of
\Fig{fig:balance}. The good correspondence indicates that these
equations can be used to investigate the role of the stresses in
generation of the mean flows. 

We emphasise, that although in the steady state, for example at the
equator, the two terms on the RHS of \Eq{eq:bbmuy} balance, these
terms are not the generators of the mean flow. They do, however,
determine the final amplitude of the flow. 
Instead, the mean flows are generated by the gradient of the vertical
stress $\qyz$ at the vertical boundaries, as can be seen from 
\Eq{eq:muytf}. This flow then slowly penetrates to the middle of
domain. Such behavior can also be clearly seen in the first panel of
\Fig{fig:uxy27}, where we show the time evolution of $\muy$. 

The generation of mean flows is straightforward at the equator, because 
the meridional stress, and hence the meridional flow vanish there.
At other latitudes the meridional stress and flow has
to be included, but it is clear that the Reynolds stresses are the
main driver of mean flows in the current setups. 

\section{Parameterization of Reynolds stresses in terms of mean--field
  hydrodynamics} 
\label{sec:Lambda}

Based on the $\Lambda$--effect theory explained in \Sec{sec:theory},
the vertical and horizontal Reynolds stresses given in
\Eqs{eq:qrp}{eq:qtp}, respectively, can be written in the simulation
domain as 
\begin{equation}
\qyz\!=\!Q_{yz}^{(\nu)}\!+\!Q_{yz}^{(\Lambda)}\!=\!-\nu_{\parallel}
\frac{\partial \muy}{\partial z}\!+\!\nu_{\parallel}V\sin \theta
\Omega,\label{eq:qqyz} 
\end{equation}
\vspace{-.5cm}
\begin{equation}
\qxy\!=\!Q_{xy}^{(\nu)}\!+\!Q_{xy}^{(\Lambda)}\!=\!\nu_{\perp}\Omega^2\sin\theta\cos
\theta 
\frac{\partial \muy}{\partial z}\!+\!\nu_{\parallel}H\cos
\theta\Omega.\label{eq:qqxy}
\end{equation}
Measuring the $\Lambda$--effect coefficients $V$ and $H$
from a single experiment is not possible, because also the turbulent
viscosities $\nupar$ and $\nuper$ are unknown.
Our strategy around this is to run another set of otherwise identical
simulations, but where the horizontal mean flows
are artificially suppressed at each time step. Therefore, the first terms in 
\Eqs{eq:qqyz}{eq:qqxy} go to zero. Then from these simulations we can directly
measure $Q^{(\Lambda)}$.
However, we need to validate this approach because the velocities can be
affected through the non--linearity of the Navier--Stokes equations.
Therefore, we perform yet another set of otherwise identical simulations,
but use periodic boundary conditions (PBC) in all directions instead of 
stress--free boundary condition (SFBC) in the vertical direction.
Then we compare the two sets of stresses obtained with these sets of
boundary conditions. 
Such a comparison of varying boundary conditions is important also in
the respect of interpreting the $\con$ dependence as depth dependence
- this approach is somewhat artificial, as we practically enforce
unrealistic BCs within the convection zone. 

As an example, we show the horizontal mean velocities for the PBC version of
\textbf{C46} at $30^\circ$ latitude in panel (A) of \Fig{fig:pbc}.
Clearly no notable mean flow is generated in this run. 
Therefore, the first term in both \Eqs{eq:qqyz}{eq:qqxy} goes to zero
similar to the cases where the mean flows are suppressed.
In panel (B) of \Fig{fig:pbc}, we show the results of the
comparison of the Reynolds stresses between PBC and SFBC cases. 
The difference caused by varying boundary conditions is confined to a
very narrow layer near the boundary. These results suggest that our
method for the separation of different effects and enforcing
artificial SFBC at different depth is valid.       

Considering \Eq{eq:qqyz}, the subtraction of the
Reynolds stresses obtained from these simulations from the total ones
gives
\EQ
Q_{yz}-Q_{yz}^{(\Lambda)}=-\nu_{\parallel} \frac{\partial
  \muy}{\partial  z}.
\label{eq:qnu}
\EN
Measuring the vertical gradient of $\muy$, the value of
$\nu_{\parallel}$ can be determined by performing an error--weighted
linear least--squares fit to \Eq{eq:qnu}. Putting the measured values
of $\nupar$ back into $Q^{(\Lambda)}$ of both \Eqs{eq:qqyz}{eq:qqxy},
we can measure $V$ and $H$ provided that $\nu_{\perp}\ll\nu_{\parallel}$.

\subsection{Properties of the diffusive and non--diffusive parts of
  Reynolds stresses}  

\begin{figure}[t!]
\begin{center}
  \includegraphics[width=\columnwidth]{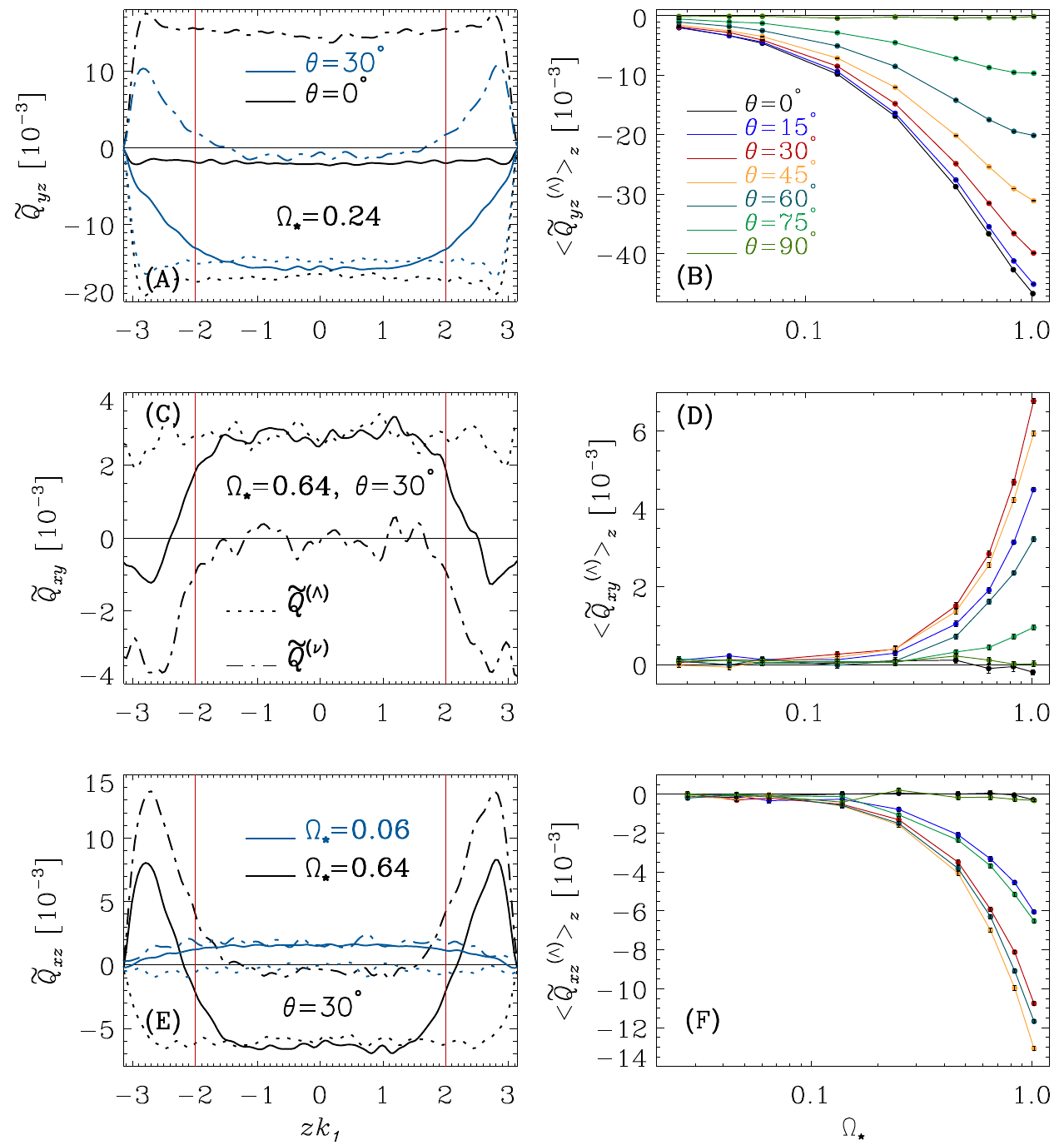}
\end{center}
\caption[]{Panels (A), (C) and (E): time--averaged diffusive and 
non--diffusive parts of the  Reynolds stresses versus vertical direction.
  The black and blue lines in panel (A) show the normalized vertical
  stresses at the equator 
  and $30^\circ$ latitude for set \textbf{C24}, respectively. In
  panel (C) the horizontal stresses are shown at $30^\circ$
  latitude for set  \textbf{C64}. The blue and black lines at the panel
  (E) show the meridional stresses for set \textbf{C06} and
 \textbf{C64} at $30^\circ$ latitude, respectively. The vertical lines
  denotes the $z$ range used for volume averages. Solid, dotted and
  dash--dotted lines show $\widetilde{Q}_{ij}$,
  $\widetilde{Q}_{ij}^{(\Lambda)}$, and $\widetilde{Q}_{ij}^{(\nu)}$,
  respectively. Panels (B), (D), and (F): volume averages, over $-2
  \le z k_1 \le 2$, of $\widetilde{Q}_{ij}^{(\Lambda)}$ versus $\con$
  at different latitudes as indicated by the legend.
}\label{fig:Qln}\end{figure}

Similar to \Sec{res:RS}, we first measure ${Q_{ij}}$ from a
non-rotating run and then subtract its mean value from corresponding
stress in other sets.  
We show the different contributions to the Reynolds stresses in \Fig{fig:Qln}. 
In the left column we show stresses from one or two simulation sets,
and in the right column we show the dependence of volume averages of
$Q_{ij}^{(\Lambda)}$ on both latitude and $\con$. 
In panel (A), we show the vertical stresses for set \textbf{C24} at
the equator and at $30^\circ$ latitude. With these results we can
explain the minimum  
of $\qyz$ at the equator: the diffusive and non--diffusive 
parts of the stresses are comparable but of opposite signs,
leading to a small negative value for the total. With
\Eq{eq:qqyz}, we see that $\nupar>0$ which in combination with
$\partial_z\muy<0$, gives $Q_{yz}^{(\nu)}>0$. Moreover, the final
negative value of $\qyz$ also shows that $Q_{yz}^{(\Lambda)}$ is
responsible for the generation of the zonal flow.  
The profile of $\qyzn$ for all other latitudes is similar to the one
at $30^\circ$, and shows that the major contribution from
the diffusive part is happening close to the boundaries at
$|zk_1|\gtrsim 2$ with positive values. Furthermore, the amplitude of
$\qyzn$ decreases towards higher latitudes (not shown). In the middle
of the domain it has negative values, which fits well with 
$\partial_z\muy>0$ that can be seen in \Fig{fig:muco}. 
The non--diffusive part of the vertical stress is always
$\qyzl<0$. Its absolute value increases from the pole towards the
equator and increases with $\con$. We also find that
$Q_{yz}^{(\Lambda)}$ is linearly dependent on $\con$ in slow rotation
regime $\con\ll 1$ in agreement with previous numerical results
\citep{PK19}.  
 \begin{figure*}[t!]
\begin{center}
\includegraphics[width=1.3\columnwidth]{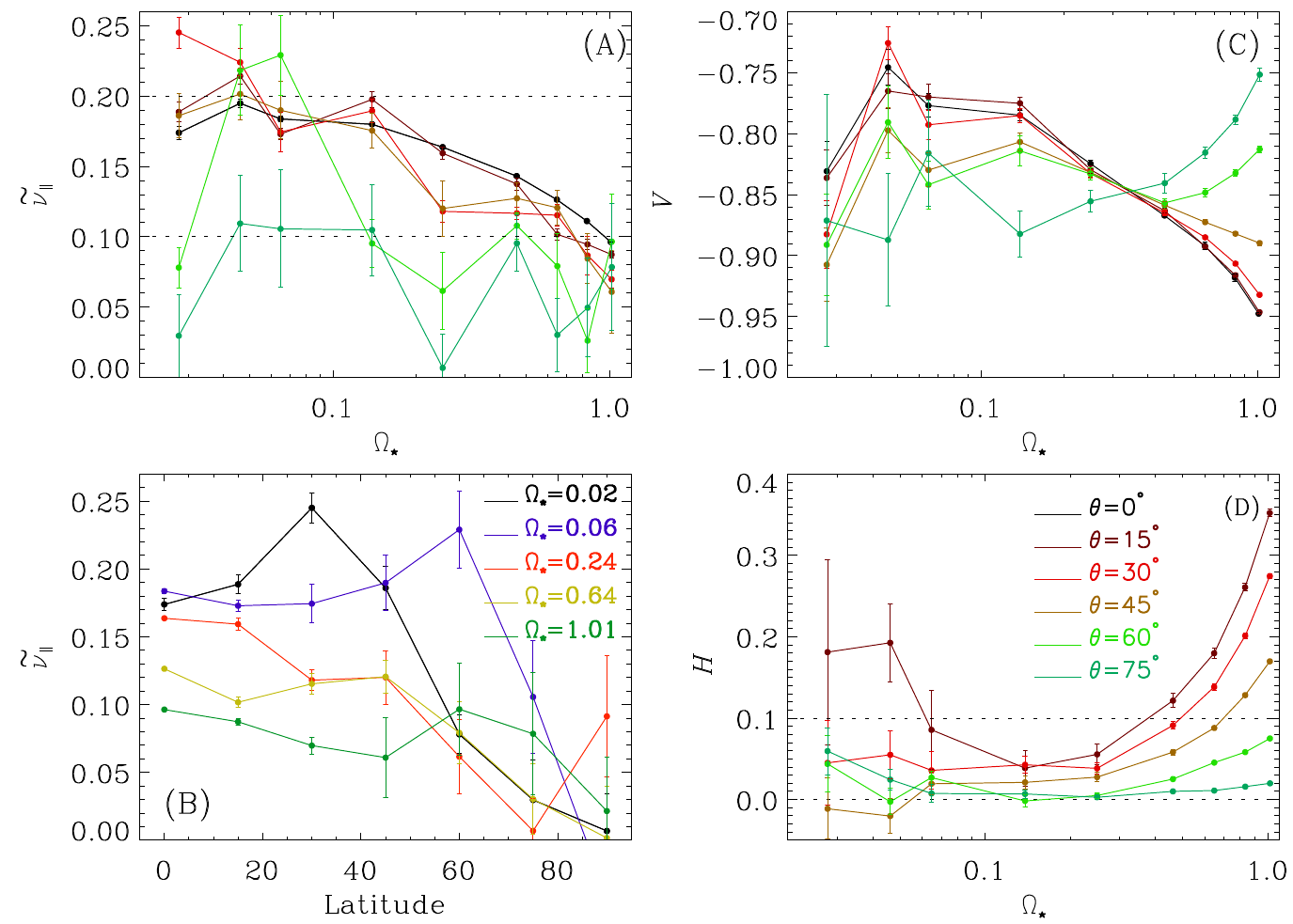}
\end{center}
\caption[]{
  Normalized turbulent viscosity $\tnupar$ and
  $\Lambda$--effect coefficients as function of $\con$ and latitude. Panels
  (A), (C) and (D) show $\tnupar$, $V$ and $H$ as a function of $\con$
  from the equator to $75^\circ$ latitude, respectively. 
  Panel (B) shows $\tnupar$ as a function of latitudes for five selected $\con$.
}\label{fig:coef}\end{figure*}

We show corresponding results for $Q_{xy}$ in panel (C) of
\Fig{fig:Qln} for $\con=0.64$. $\qxyl$ has positive values in the
whole domain while $\qxyn$ is almost zero in the middle of the domain
and its contribution to $\qxy$ confined to the boundaries at
$|zk_1|\gtrsim 2$. This also shows that $\qxyn$ is the main
contributor to the negative values of $\qxy$ close to the boundaries
shown in \Fig{fig:allq}\,(C). The volume averaged values of $\qxyl$,
excluding the boundaries, are shown in \Fig{fig:Qln}\,(D) as a
function of both $\con$ and latitude.
Its value is almost zero both at the equator and at the pole. It is 
significantly non--zero above $\con> 0.24$ and increases with
increasing $\con$ independent of latitude. Independent of $\con$, it
has maximum value at $30^\circ$ latitude. 
We note here that the amplitude of $\qxyl$ is also significantly
smaller than $\qyzl$. The measured profile of $\qxyl$ is almost
identical to the one obtained by \cite{Kap19}. 

Our results for $Q_{xz}$ are shown in \Fig{fig:Qln}. At low $\con$,
there is almost no contribution of $\tqxzl$ to the total stresses. For
$\con>0.15$, the contribution of $\tqxzn$ disappears in the middle of
the domain but maintains its positive value close to the
boundaries. This can be seen in panel (E) where we show $\tqxz$ for
low and high $\con$ for the sake of comparison. In panel (F), we show
volume averages of $\tqxzl$ at all $\con$ 
and latitudes. The value of $\tqxzl$ is almost zero at the equator and at the
pole. In other latitudes, its absolute value increases with increasing $\con$.
It has always negative values independent of both $\con$ and latitude. 
These results are in agreement with those of \cite{Kap19} in the slow
rotation regime. 

\subsection{Measuring turbulent viscosity}
\label{sec:nu}

The diagonal turbulent viscosity $\nupar$, normalized by $\ell\urms$, and its
dependence on both $\con$ and latitude is shown in panels (A) 
and (B) of \Fig{fig:coef}, respectively. Apart from the highest latitudes 
where measurements are unreliable, the turbulent viscosity decreases 
monotonically as a function of $\con$ such that for the largest $\con$, 
corresponding to bottom of the NSSL, its value has decreased by roughly a 
factor of two. 
The method used here to measure turbulent viscosity relies on the presence of 
mean flows. As these diminish toward high latitudes it is very difficult to 
obtain reliable estimates of $\nupar$ near the pole. We note that the
measurements of $\tnupar$ also suffer from
numerical noise at $\con<0.1$ at low latitudes. In particular, we 
think that the latitudinal dependence of $\tnupar$ for $\theta\lesssim60\degr$, 
shown in panel (B), is not reliable. According to the results at lower
latitudes, we conclude that the latitude dependence is weak in
comparison to the rotational dependence. Hence, we consider the
profile of $\nupar$ at the equator applicable for other latitudes
which is measured with high confidence and use it for measuring $V$
and $H$ at other latitudes. The ratio of turbulent to kinematic
viscosity is $\nupar/\nu\sim 10$-$20$, as expected for the fluid
Reynolds numbers in the current simulations. 

\cite{KRBK20} measured turbulent viscosity by imposing a weak
sinusoidal shear flow in a non--rotating isotropically forced
turbulent medium in a Cartesian geometry and measured the response of
the system. The response is an off--diagonal Reynolds stress 
that is assumed to be proportional to the imposed shear flow according to the 
Boussinesq ansatz. They defined a shear number Sh as
\EQ
{\rm Sh}=\frac{U_0k_U}{\urms k_f},
\EN
where $U_0$ is the amplitude of the flow and $k_U$ is the wavenumber
of the imposed sinusoidal shear. To obtain Sh for the shear flows
generated at the equator in the present simulations, we set
$k_U=k_1/2$ and $U_0/\urms={\rm max}(\tmuy)$. We consider only the
slow rotation regime where $\con<0.1$, corresponding to
sets~\textbf{C02}, \textbf{C04} and \textbf{C06}, where ${\rm
  Sh}=0.002,0.004$ and $0.006$ which is within the range of {\rm Sh}
values used in \cite{KRBK20}. We normalized our $\nupar$ with the same
normalization factor as in \cite{KRBK20}, that is
$\nu_{t0}=\urms/3k_f$; see their Section~2.2. This differs from the
currently used normalization by a factor of $6\pi$ such that the
current normalized values, for example in Fig.~\ref{fig:coef}, are
smaller than theirs by this factor. Using their normalization, we
obtain values of $\nut/\nu_{t0}\approx 3.5\ldots 3.8$ which are roughly
twice larger than in \cite{KRBK20}. The difference is likely caused by
the presence of strong anisotropy of turbulence in our simulations due
to the forcing and the rotation that were absent in the study of
\cite{KRBK20}. 

\begin{figure}[t!]
\begin{center}
\includegraphics[width=\columnwidth]{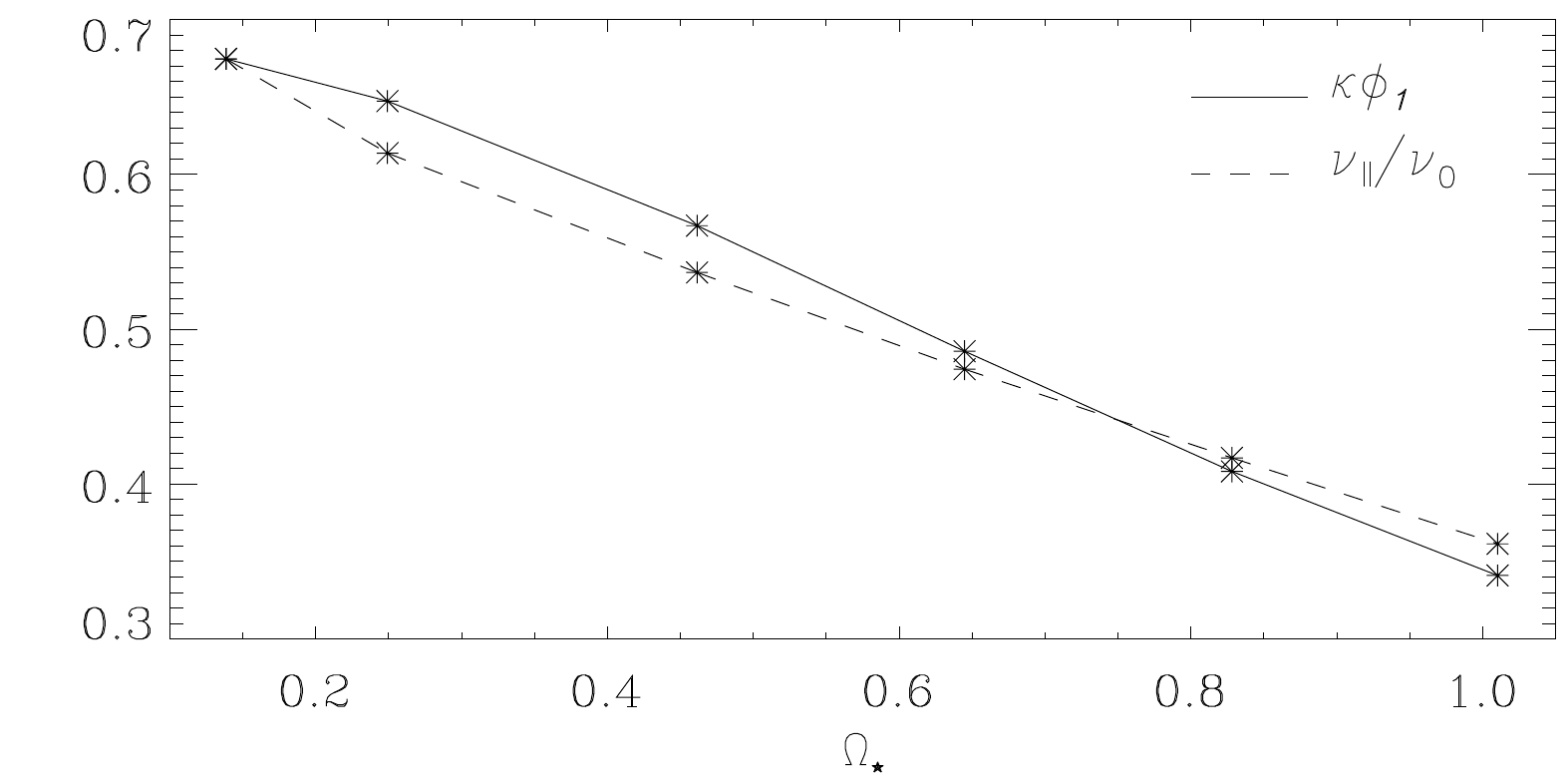}
\end{center}\caption{
Comparison of obtained turbulent viscosity with analytical result of
KPR94. The solid and dashed line show the normalized turbulent
viscosity and rescaled analytical expression $\kappa\phi_1$,
respectively. 
}\label{fig:phi1}\end{figure}

We also compare the profile of $\nupar$ with an analytical expression for the rotation
dependence of the viscosity obtained under SOCA by \citet[][hereafter
  KPR94]{KPR94}.  We consider the first term in Equation (34) of their
work which is relevant to our simulations in which $\nupar=\nu_0
\phi_1(\con)$, where $\nu_0=4/15\ell\urms$ is the turbulent viscosity
obtained for the isotropic non--rotating case, and where $\phi_1$ is a
function of $\con$ given in the Appendix of KPR94. We scale the
analytical result by a factor of $\kappa=0.68$ to make it comparable
with our numerical result. In \Fig{fig:phi1}, we show the result of
this comparison. This result shows that apart from $\kappa$ factor the
rotation dependence is in fair agreement between the theory and
numerical simulations. 
 
Considering the off--diagonal turbulent viscosity $\nuper$, we failed
measuring it as both terms constituting it, $\qxyn$ and
$\Omega^2\sin\theta\cos \theta \partial \muy/\partial z$, are too
small, and the measurement error in the former is large. 

\subsection{Measurements of the vertical $\Lambda$--effect coefficient}

We measure the vertical $\Lambda$--coefficient by substituting the volume 
averages of $\qyz^{(\Lambda)}$ shown in panel (A) of \Fig{fig:Qln} and
$\nupar$ at the equator using 
\EQ
V=\frac{\qyz^{(\Lambda)}}{\nupar \sin\theta \Omega_0}.
\EN
Our results are shown in panel (C) of \Fig{fig:coef}. 
The absolute value of $V$ is about $0.75$ and gradually increases
to $\approx 0.95$ for latitudes $\leq 45^\circ$. However, the value of
$V$ at the lowest 
$\con$ are smaller at all latitudes, but they have large error bars. 
In contrast to low latitudes, the absolute values of $V$ at $60^\circ$
and $75^\circ$ decrease for $\con>0.3$. Considering the large errors
in the measurements at low $\con$, we might consider $V$ being roughly
constant for $\con\leq 0.15$ independent of latitude, but it shows
strong latitudinal and rotational dependency for $\con > 0.15$. This
means that considering only the first term $V^{(0)}$ in \Eq{eq:v} in
the NSSL condition is not enough as it is assumed by the theoretical
model by KR05 explained in \Sec{sec:theory}. Moreover, the increase of
$V$ towards higher $\con$ at low latitudes is in contrast with the
decrease predicted by KR05 model. 
The same applies to the results of \cite{Kap19} who did not consider that
$\nut=\nut(\Omega_\star)$.

\subsection{Measurements of the horizontal $\Lambda$--effect coefficient}

We measure the horizontal $\Lambda$--effect coefficient similarly to the
vertical one using  
\EQ
H=\frac{\qxy^{(\Lambda)}}{\nupar \cos\theta \Omega_0}.
\EN
The results are shown in panel (D) of \Fig{fig:coef}. The values
of $H$ are always positive independent of $\con$ and latitude. Its
values are one order of magnitude smaller than $V$ up to
$\con=0.6$, above which $H$ begins to increase
at latitudes $<45^\circ$. We also note that its value is zero at the
equator and at the pole. $H$ the largest at $15^\circ$ and
decreases gradually towards higher latitudes.
These results show that $H$ does not play any role close to
the surface in transporting the angular momentum which validates the
assumption applied in the NSSL model by KR05.

\section{Conclusions}

We applied an alternative approach to MF and GDNS, namely running 
direct numerical simulations of forced turbulence in local boxes, 
to primarily find out if the assumptions and approximations applied in
MF theory to explain the formation of the NSSL are valid. In contrast
to GDNS, we could isolate and study the role and contribution of the
Reynolds stresses in the rotational regime relevant for the NSSL. 
Additionally, we were able to measure the turbulent viscosity. Our
results show that applying the three required conditions, explained in
\Sec{sec:theory}, that are necessary to generate the NSSL in the RK05
model are insufficient. In particular, the meridional component of the
Reynolds stress cannot be ignored. However, our results are in
accordance with $Q_{xy}\rightarrow 0$ in the upper part of the NSSL,
whereas $\qxy$ obtains small but non--zero values close to the bottom
of the NSSL in agreement with the theoretical predictions. Regarding
the vertical Reynolds stress, its role in transporting the  
angular momentum radially inward is in agreement with theory. However,
its profile differs from that predicted by theory. 
In particular, it was assumed in \cite{LK13} and \cite{LK16} that only
the term $V^{(0)}$ survives in the expansion of $V$ in the
NSSL. However, our results indicate that higher order 
terms in the expansion of $V$ need to be considered. Moreover, it is
also expected from theory that the vertical transport of angular
momentum decreases by increasing $\con$ independent of latitude, but
our results show that this expectation is fulfilled only at high
latitudes. We also note here that the rotational quenching of the
turbulent viscosity, $\nupar$, adds another degree of complexity to
the problem which was not considered previously in the models of NSSL.
This behavior, however, from a theoretical MF prediction is in good
qualitative agreement with our results \citep{KPR94}.

Although these local box simulations have a moderate value of ${\rm
  Re}\approx13$, and there is no connection between different
latitudes, our results are largely consistent with the stresses and
mean flows obtained in GDNS. 
On the other hand, the theoretical works used SOCA which should be
valid at Reynolds or Strouhal numbers of up to unity, which is in the
vicinity of the parameter regime of the current models. Hence, it is
not surprising that we find a relatively good match in between the
measured turbulent viscosity and the one predicted by SOCA. 

Concerning the fact that $Q_{xz}$ cannot be disregarded in the NSSL,
its role can be further investigated in more realistic setup using
spherical geometry where the artifact of discontinuity between
latitudes can be removed. We also note here that in this work we
consider only a single modest Reynolds number and one forcing scale,
the effects of which need to be explored with wider parameter
studies. The other important physics that need to be investigated  
are the effect of stratification, compressibility and magnetic fields,
and comparing with previous studies that have studied these in
turbulent convection, namely \cite{PTBNS93}, \cite{Chan2001}, and
\cite{KKT04}.

It is worthwhile to note here that a set of companion laboratory
experiments is being proposed to test several aspects of our model. In
these experiments, a rotating water--filled apparatus will be used to
simulate regions of finite latitudinal extent, including
$\beta$--plane effects, and forcing will be introduced by pump--driven
nozzles at the boundaries \citep{EH19}. Relative variation of the
system rotation rate and the nozzle exit velocity will allow both the
$\Omega_\star > 1 $ and $\Omega_\star < 1$ regimes to be explored. The
forcing scale length and isotropy will be changed by opening/closing
nozzles and altering the nozzle shapes and orientations. Time resolved
measurements of the components of the flow velocity will allow the
mean flows and stresses to be computed and compared with numerical
results and theoretical models. Despite that the details of the
forcing and the fluid boundary conditions will be different in the
experiment than in the present simulations, it is expected that
meaningful results will be obtained as the rotation rate of the system
is varied and the experimental data is analyzed to look for signatures
of the $\Lambda$--effect. 
 
\begin{acknowledgements}
  The simulations have been carried out on supercomputers at GWDG and
  on the Max Planck supercomputer at RZG in Garching. 
This project has received funding from the European Research Council (ERC) 
under the European Union's Horizon 2020 research and innovation 
programme (project "UniSDyn", grant agreement n:o 818665).
A. Barekat acknowledges funding by the Max-Planck/Princeton Center for
Plasma Physics. PJK acknowledges financial support from the Deutsche
Forschungsgemeinschaft (DFG) Heisenberg programme (grant No.\ KA 4825/2-1).
\end{acknowledgements}

\bibliographystyle{aa}
\bibliography{all}

\newcommand{\noop}[1]{}
\begin{thebibliography}{50}
\expandafter\ifx\csname natexlab\endcsname\relax\def\natexlab#1{#1}\fi

\bibitem[{{Barekat} {et~al.}(2014){Barekat}, {Schou}, \& {Gizon}}]{BSG14}
{Barekat}, A., {Schou}, J., \& {Gizon}, L. 2014, \aap, 570, L12

\bibitem[{{Barekat} {et~al.}(2016){Barekat}, {Schou}, \& {Gizon}}]{BSG16}
{Barekat}, A., {Schou}, J., \& {Gizon}, L. 2016, \aap, 595, A8

\bibitem[{{Brandenburg}(2001)}]{AB01}
{Brandenburg}, A. 2001, \apj, 550, 824

\bibitem[{{Brandenburg}(2005)}]{AB05}
{Brandenburg}, A. 2005, \apj, 625, 539

\bibitem[{{Brandenburg} {et~al.}(2020){Brandenburg}, {Johansen}, {Bourdin},
  {Dobler}, {Lyra}, {Rheinhardt}, {Bingert}, {Haugen}, {Mee}, {Gent},
  {Babkovskaia}, {Yang}, {Heinemann}, {Dintrans}, {Mitra}, {Candelaresi},
  {Warnecke}, {K{\"a}pyl{\"a}}, {Schreiber}, {Chatterjee}, {K{\"a}pyl{\"a}},
  {Li}, {Kr{\"u}ger}, {Aarnes}, {Sarson}, {Oishi}, {Schober}, {Plasson},
  {Sandin}, {Karchniwy}, {Rodrigues}, {Hubbard}, {Guerrero}, {Snodin},
  {Losada}, {Pekkil{\"a}}, \& {Qian}}]{PC20}
{Brandenburg}, A., {Johansen}, A., {Bourdin}, P.~A., {et~al.} 2020, arXiv
  e-prints, arXiv:2009.08231

\bibitem[{{Burin} {et~al.}(2019){Burin}, {Caspary}, {Edlund}, {Ezeta},
  {Gilson}, {Ji}, {McNulty}, {Squire}, \& {Tynan}}]{EH19}
{Burin}, M.~J., {Caspary}, K.~J., {Edlund}, E.~M., {et~al.} 2019, \pre, 99,
  023108

\bibitem[{{Chan}(2001)}]{Chan2001}
{Chan}, K.~L. 2001, \apj, 548, 1102

\bibitem[{{Chandrasekhar}(1961)}]{Ch61}
{Chandrasekhar}, S. 1961, {Hydrodynamic and hydromagnetic stability}

\bibitem[{{Duvall}(1979)}]{TD:79}
{Duvall}, T.~L., J. 1979, \solphys, 63, 3

\bibitem[{{Foukal} \& {Jokipii}(1975)}]{FJ75}
{Foukal}, P. \& {Jokipii}, J.~R. 1975, Astrophys. Journal Letters, 199, L71

\bibitem[{{Gilman}(1977)}]{Gi77}
{Gilman}, P.~A. 1977, Geophys. Astrophys. Fluid Dynam., 8, 93

\bibitem[{{Gilman}(1983)}]{Gi83}
{Gilman}, P.~A. 1983, \apjs, 53, 243

\bibitem[{{Gilman} \& {Howard}(1984)}]{GH84}
{Gilman}, P.~A. \& {Howard}, R. 1984, \solphys, 93, 171

\bibitem[{{Glatzmaier}(1985)}]{Gl85}
{Glatzmaier}, G.~A. 1985, \apj, 291, 300

\bibitem[{{Guerrero} {et~al.}(2016){Guerrero}, {Smolarkiewicz}, {de Gouveia Dal
  Pino}, {Kosovichev}, \& {Mansour}}]{GSdGDPKM16}
{Guerrero}, G., {Smolarkiewicz}, P.~K., {de Gouveia Dal Pino}, E.~M.,
  {Kosovichev}, A.~G., \& {Mansour}, N.~N. 2016, \apj, 819, 104

\bibitem[{{Guerrero} {et~al.}(2013){Guerrero}, {Smolarkiewicz}, {Kosovichev},
  \& {Mansour}}]{2013ApJ...779..176G}
{Guerrero}, G., {Smolarkiewicz}, P.~K., {Kosovichev}, A.~G., \& {Mansour},
  N.~N. 2013, \apj, 779, 176

\bibitem[{{Gunderson} \& {Bhattacharjee}(2019)}]{GB19}
{Gunderson}, L.~M. \& {Bhattacharjee}, A. 2019, \apj, 870, 47

\bibitem[{{Hanasoge} {et~al.}(2016){Hanasoge}, {Gizon}, \&
  {Sreenivasan}}]{HGS16}
{Hanasoge}, S., {Gizon}, L., \& {Sreenivasan}, K.~R. 2016, Annual Review of
  Fluid Mechanics, 48, 191

\bibitem[{{Hathaway}(1996)}]{DH96}
{Hathaway}, D.~H. 1996, \apj, 460, 1027

\bibitem[{{Hathaway} \& {Upton}(2014)}]{HU14}
{Hathaway}, D.~H. \& {Upton}, L. 2014, Journal of Geophysical Research (Space
  Physics), 119, 3316

\bibitem[{{Hotta} {et~al.}(2015){Hotta}, {Rempel}, \& {Yokoyama}}]{HRY15}
{Hotta}, H., {Rempel}, M., \& {Yokoyama}, T. 2015, \apj, 798, 51

\bibitem[{{K{\"a}pyl{\"a}}(2019{\natexlab{a}})}]{PK19}
{K{\"a}pyl{\"a}}, P.~J. 2019{\natexlab{a}}, Astronomische Nachrichten, 340, 744

\bibitem[{{K{\"a}pyl{\"a}}(2019{\natexlab{b}})}]{Kap19}
{K{\"a}pyl{\"a}}, P.~J. 2019{\natexlab{b}}, \aap, 622, A195

\bibitem[{{K{\"a}pyl{\"a}} \& {Brandenburg}(2008)}]{KB08}
{K{\"a}pyl{\"a}}, P.~J. \& {Brandenburg}, A. 2008, \aap, 488, 9

\bibitem[{{K{\"a}pyl{\"a}} {et~al.}(2006){K{\"a}pyl{\"a}}, {Korpi}, \&
  {Tuominen}}]{KKT06}
{K{\"a}pyl{\"a}}, P.~J., {Korpi}, M.~J., \& {Tuominen}. 2006, Astronomische
  Nachrichten, 327, 884

\bibitem[{{K{\"a}pyl{\"a}} {et~al.}(2004){K{\"a}pyl{\"a}}, {Korpi}, \&
  {Tuominen}}]{KKT04}
{K{\"a}pyl{\"a}}, P.~J., {Korpi}, M.~J., \& {Tuominen}, I. 2004, \aap, 422, 793

\bibitem[{{K{\"a}pyl{\"a}} {et~al.}(2011{\natexlab{a}}){K{\"a}pyl{\"a}},
  {Mantere}, \& {Brandenburg}}]{KMB11}
{K{\"a}pyl{\"a}}, P.~J., {Mantere}, M.~J., \& {Brandenburg}, A.
  2011{\natexlab{a}}, Astron. Nachr., 332, 883

\bibitem[{{K{\"a}pyl{\"a}} {et~al.}(2011{\natexlab{b}}){K{\"a}pyl{\"a}},
  {Mantere}, {Guerrero}, {Brandenburg}, \& {Chatterjee}}]{KMGB11}
{K{\"a}pyl{\"a}}, P.~J., {Mantere}, M.~J., {Guerrero}, G., {Brandenburg}, A.,
  \& {Chatterjee}, P. 2011{\natexlab{b}}, \aap, 531, A162

\bibitem[{{K{\"a}pyl{\"a}} {et~al.}(2020){K{\"a}pyl{\"a}}, {Rheinhardt},
  {Brandenburg}, \& {K{\"a}pyl{\"a}}}]{KRBK20}
{K{\"a}pyl{\"a}}, P.~J., {Rheinhardt}, M., {Brandenburg}, A., \&
  {K{\"a}pyl{\"a}}, M.~J. 2020, \aap, 636, A93

\bibitem[{{Kitchatinov}(2013)}]{LK13}
{Kitchatinov}, L.~L. 2013, in IAU Symposium, Vol. 294, IAU Symposium, ed. A.~G.
  {Kosovichev}, E.~{de Gouveia Dal Pino}, \& Y.~{Yan}, 399--410

\bibitem[{{Kitchatinov}(2016)}]{LK16}
{Kitchatinov}, L.~L. 2016, Astronomy Letters, 42, 339

\bibitem[{{Kitchatinov} {et~al.}(1994){Kitchatinov}, {Pipin}, \&
  {Ruediger}}]{KPR94}
{Kitchatinov}, L.~L., {Pipin}, V.~V., \& {Ruediger}, G. 1994, Astronomische
  Nachrichten, 315, 157

\bibitem[{{Kitchatinov} \& {R{\"u}diger}(2005)}]{LKGR05}
{Kitchatinov}, L.~L. \& {R{\"u}diger}, G. 2005, Astronomische Nachrichten, 326,
  379

\bibitem[{{Krause} \& {R{\"a}dler}(1980)}]{1980opp..bookR....K}
{Krause}, F. \& {R{\"a}dler}, K.~H. 1980, {Mean-field magnetohydrodynamics and
  dynamo theory}

\bibitem[{{Lebedinski}(1941)}]{Leb41}
{Lebedinski}, A.~I. 1941, Astron. Zh., 18, 10

\bibitem[{{Matilsky} {et~al.}(2019){Matilsky}, {Hindman}, \& {Toomre}}]{MT19}
{Matilsky}, L.~I., {Hindman}, B.~W., \& {Toomre}, J. 2019, \apj, 871, 217

\bibitem[{{Miesch} \& {Hindman}(2011)}]{MH11}
{Miesch}, M.~S. \& {Hindman}, B.~W. 2011, \apj, 743, 79

\bibitem[{{Parker}(1955)}]{EP95}
{Parker}, E.~N. 1955, \apj, 122, 293

\bibitem[{{Pulkkinen} \& {Tuominen}(1998)}]{PT98}
{Pulkkinen}, P. \& {Tuominen}, I. 1998, \aap, 332, 755

\bibitem[{{Pulkkinen} {et~al.}(1993){Pulkkinen}, {Tuominen}, {Brandenburg},
  {Nordlund}, \& {Stein}}]{PTBNS93}
{Pulkkinen}, P., {Tuominen}, I., {Brandenburg}, A., {Nordlund}, A., \& {Stein},
  R.~F. 1993, \aap, 267, 265

\bibitem[{{Robinson} \& {Chan}(2001)}]{RC01}
{Robinson}, F.~J. \& {Chan}, K.~L. 2001, \mnras, 321, 723

\bibitem[{{R{\"u}diger}(1980)}]{GR80}
{R{\"u}diger}, G. 1980, Geophys. Astrophys. Fluid Dynam., 16, 239

\bibitem[{{R{\"u}diger}(1989)}]{GR89}
{R{\"u}diger}, G. 1989, Differential rotation and stellar convection. Sun and
  the solar stars (Berlin: Akademie Verlag, 1989)

\bibitem[{{R{\"u}diger} {et~al.}(2019){R{\"u}diger}, {K{\"u}ker},
  {K{\"a}pyl{\"a}}, \& {Strassmeier}}]{RKKS19}
{R{\"u}diger}, G., {K{\"u}ker}, M., {K{\"a}pyl{\"a}}, P.~J., \& {Strassmeier},
  K.~G. 2019, \aap, 630, A109

\bibitem[{{Schou} {et~al.}(1998){Schou}, {Antia}, {Basu}, {Bogart}, {Bush},
  {Chitre}, {Christensen-Dalsgaard}, {Di Mauro}, {Dziembowski}, {Eff-Darwich},
  {Gough}, {Haber}, {Hoeksema}, {Howe}, {Korzennik}, {Kosovichev}, {Larsen},
  {Pijpers}, {Scherrer}, {Sekii}, {Tarbell}, {Title}, {Thompson}, \&
  {Toomre}}]{JS98}
{Schou}, J., {Antia}, H.~M., {Basu}, S., {et~al.} 1998, \apj, 505, 390

\bibitem[{{Stix}(2002)}]{Stix:2000}
{Stix}, M. 2002, {The sun: an introduction}

\bibitem[{{Thompson} {et~al.}(1996){Thompson}, {Toomre}, {Anderson}, {Antia},
  {Berthomieu}, {Burtonclay}, {Chitre}, {Christensen-Dalsgaard}, {Corbard}, {De
  Rosa}, {Genovese}, {Gough}, {Haber}, {Harvey}, {Hill}, {Howe}, {Korzennik},
  {Leibacher}, {Pijpers}, {Provost}, {Rhodes}, {Schou}, {Sekii}, {Stark}, \&
  {Wilson}}]{MT96}
{Thompson}, M.~J., {Toomre}, J., {Anderson}, E.~R., {et~al.} 1996, Science,
  272, 1300

\bibitem[{{Ward}(1965)}]{FW65}
{Ward}, F. 1965, \apj, 141, 534

\bibitem[{{Warnecke} {et~al.}(2016){Warnecke}, {K{\"a}pyl{\"a}},
  {K{\"a}pyl{\"a}}, \& {Brandenburg}}]{WKKB16}
{Warnecke}, J., {K{\"a}pyl{\"a}}, P.~J., {K{\"a}pyl{\"a}}, M.~J., \&
  {Brandenburg}, A. 2016, \aap, 596, A115

\bibitem[{{Yoshimura}(1975)}]{HY75}
{Yoshimura}, H. 1975, \apj, 201, 740

\end{thebibliography}

\end{document}